\documentclass[a4paper,11pt]{article}
\usepackage{jheppub} 
\usepackage{lineno}
\usepackage{tensor}
\usepackage{bm,bbm,dsfont}
\usepackage{tikz-feynman}
\usepackage{appendix}
\usepackage{mathtools}
\usepackage{amsfonts}                  
\usepackage{subcaption} 
\usepackage{braket}
\usepackage[T1]{fontenc}      
\usepackage{ae}
\usepackage{cleveref}
\usepackage{subfiles}
\usepackage{slashed}
\usepackage{lastpage}

\newcommand{\Tr}{\ensuremath{\text{Tr}}}
\newcommand{\STr}{\ensuremath{\text{STr}}}

\newcommand{\lrb}[1]{\left(#1\right)}
\newcommand{\lrmb}[1]{\left[#1\right]}

\newcommand{\rar}{\rightarrow}

\def \half {{1\over 2}}

\def \pd {\partial}
\def\pdr{\pd_\rho}
\newcommand{\pdbr}[1]{\pdr\lrb{\rho^3\pdr #1}}

\newcommand{\dmq}{\ensuremath{\D m^2}}
\newcommand{\dmqp}{\ensuremath{\D m^{2\prime}}}

\def\gsim{\raise0.3ex\hbox{$\;>$\kern-0.75em\raise-1.1ex\hbox{$\sim\;$}}}
\def\lsim{\raise0.3ex\hbox{$\;<$\kern-0.75em\raise-1.1ex\hbox{$\sim\;$}}}

\def \gsq {g_5^2}

\newcommand{\invrud}{\lrb{{1\over r_1^4}+{1\over r_6^4}}}

\newcommand{\lun}{\ensuremath{L_1}}

\newcommand{\lpn}{\ensuremath{L_p}}
\newcommand{\lnn}{\ensuremath{L_m}}
\newcommand{\ldn}{\ensuremath{L_6}}

\newcommand{\bv}[1]{\Big|_{#1}}
\newcommand{\mcl}[1]{\ensuremath{\mathcal{#1}}}

\def\a  {\alpha}                
       \def\d  {\delta}        \def\D  {\Delta}
        
      \def\L  {\Lambda}       
          \def\s  {\sigma}

\title{\Large Holography for Sp(2$N_c$) Gauge Dynamics: from Composite Higgs to Technicolour}

\author[a]{Johanna Erdmenger,}
\author[b]{Nick Evans,}
\author[a,1]{Yang Liu,\note{Corresponding author.}}
\author[a]{Werner Porod {}}
\affiliation[a]{Institute for Theoretical Physics and Astrophysics, Julius-Maximilians-Universit\"at W\"urzburg,\\
97074 W\"urzburg, Germany}
\affiliation[b]{School of Physics \& Astronomy and STAG Research Centre, University of Southampton,\\
Highfield, Southampton SO17 1BJ, UK}

\emailAdd{erdmenger@physik.uni-wuerzburg.de}\emailAdd{n.j.evans@soton.ac.uk}
\emailAdd{yang.liu@physik.uni-wuerzburg.de}
\emailAdd{porod@physik.uni-wuerzburg.de} 

\numberwithin{equation}{section} 

\abstract{\noindent We study Sp(2$N_c$) gauge dynamics with two Dirac fermion flavours in the fundamental representation. These strongly coupled systems underlie some composite Higgs models with a global symmetry breaking pattern SU(4)$\rightarrow$Sp(4), leading to a light quartet of pseudo-Goldstone bosons that can play the role of the Higgs. Including four-fermion interactions can rotate the vacuum to a technicolour breaking pattern. Using gauge/gravity duality, we study the underlying gauge dynamics.
Our model incorporates the $N_c$-specific running of the fermion anomalous dimension and can thus distinguish between specific gauge groups. We determine the bound-state spectrum of the UV SU(4)$\rightarrow$Sp(4) symmetry breaking model, also including fermion masses and mass splitting,  and display the $N_c$ dependence. The inclusion of a four-fermion interaction shows the emergence of three Goldstone bosons  on the path to technicolour dynamics. }

\keywords{Gauge/Gravity Duality, AdS/Yang-Mills, Composite Higgs, Technicolour.}

\begin{document}
\maketitle
\flushbottom

\section{Introduction}

Holography \cite{Maldacena:1997re, Witten:1998qj} is now a well-developed technique to describe strongly coupled gauge theories 
using a dual weak coupling gravity description. Although it was first understood for the highly supersymmetric ${\cal N}=4$ Super-Yang Mills (SYM) theory, it has since been expanded to describe a wider range of theories including with broken conformal and supersymmetry \cite{Freedman:1999gk,Pilch:2000ue,Pilch:2000fu,Babington:2002qt} and with added fermion matter content \cite{Karch:2002sh,Kruczenski:2003be, Erdmenger:2007cm}. More generically, the tool kit for the construction of dual theories can be used as a phenomenological tool to study any strongly coupled system in a bottom up fashion \cite{Erlich:2005qh, DaRold:2005mxj}.

In recent years we have been using the holographic technique to study a range of non-supersymmetric gauge theories with interesting dynamics \cite{Alho:2013dka, Erdmenger:2014fxa, Clemens:2017luw,Belyaev:2018jse,Erdmenger:2020lvq,Erdmenger:2020flu,Erdmenger:2023hkl}. Theories studied include QCD \cite{Erdmenger:2020flu,Erdmenger:2023hkl}, theories with strong Nambu-Jona-Lasinio (NJL) four-fermion interactions \cite{Evans:2016yas,Clemens:2017udk}, technicolour theories including with walking dynamics \cite{Alho:2013dka, Erdmenger:2014fxa, Clemens:2017luw,Belyaev:2018jse} and theories with multiple fermion representations \cite{Erdmenger:2020lvq,Erdmenger:2020flu}. Related holographic work includes \cite{Gursoy:2007cb,Gursoy:2007er,Jarvinen:2015ofa} for QCD,  and \cite{Elander:2023aow,Elander:2020nyd,Elander:2021bmt,Espriu:2020hae,Dillon:2018wye,Espriu:2017mlq,Croon:2015wba,Elander:2021kxk} for composite Higgs models \cite{Kaplan:1983fs,Kaplan:1983sm,Banks:1984gj,Terazawa:1976xx}.

In this paper we turn our attention to a particular set of gauge theories with Sp(2$N_c$) gauge group and two Dirac fermion flavours in the fundamental representation \cite{Ryttov:2008xe, Cacciapaglia:2014uja,Cacciapaglia:2020kgq}. The theory is interesting because, if it breaks chiral symmetry like QCD, then the global symmetry breaking pattern on the four pseudo-real Weyl fermions is expected to be SU(4)$\rightarrow$Sp(4).
In the large $N_c$ limit where the U(1)$_A$ anomaly is absent,  this is enhanced to U(4)$\rightarrow$Sp(4). The five (six) broken generators lead to a set of Goldstone bosons that contain a quadruplet which transforms under an SU(2)$\times$SU(2) subgroup. If this symmetry group is regarded as the chiral group of the standard model, where SU(2)$_L$ is gauged, then the quadruplet is a candidate for a composite Higgs boson \cite{Arkani-Hamed:2002ikv}. 
The low-energy effective potential that can lead to electroweak symmetry breaking is driven by loops in the low-energy theory including electroweak gauge bosons and the top quark \cite{Arkani-Hamed:2002ikv,Golterman:2015zwa} --- that mechanism is not what we study here. 
We concentrate on the high scale gauge dynamics that causes the composite state to emerge and the spectrum of other composite states of the gauge theory. 
The fifth Goldstone state has been proposed as a dark matter candidate \cite{Ryttov:2008xe,Pomper:2024otb}.
 See also \cite{Hochberg:2014kqa,Kulkarni:2022bvh,Zierler:2022uez,Kondo:2022lgg} for  models where the gauge theory we consider is entirely in the dark sector.

Another interesting aspect of the theory is that if one includes NJL operators one can bias a vacuum that explicitly breaks the electroweak chiral symmetry via a technicolour-like symmetry breaking pattern \cite{Farhi:1980xs}. Here there is a vacuum alignment problem between the composite Higgs vacuum and the technicolor like vacuum which has previously been studied at the level of the chiral Lagrangian for the theory \cite{Cacciapaglia:2014uja,Cacciapaglia:2020kgq}. For a similar construction for SO(6)/SO(5) see \cite{Barnard:2013zea}.

The holographic model we have developed \cite{Alho:2013dka,Erdmenger:2020flu,Erdmenger:2023hkl} is based on the D3/probe D7 system \cite{Karch:2002sh,Kruczenski:2003be, Erdmenger:2007cm} and in particular the Dirac Born Infeld (DBI) action for the probe branes. That theory is a rigorous top-down construction of an ${\cal N}=2$ SYM theory with fermions in the fundamental representation. That model has been shown to incorporate chiral symmetry breaking when supersymmetry is broken \cite{Babington:2003vm,Filev:2007gb}. The mechanism is that the background geometry induces a radially dependent mass term for the scalar field dual to the fermion condensate in the DBI action of the D7 brane \cite{Alvares:2012kr}. When the mass squared violates the Breitenlohner-Freedman bound (BF bound) \cite{Breitenlohner:1982jf} chiral condensation is triggered. This leads to the natural idea that one can include any given gauge dynamics by correctly encoding the running anomalous dimensions of the theory \cite{Alho:2013dka}. The DBI action then turns that input into a prediction for the mesonic spectrum of the theory. Here we use the perturbative running from the gauge theory to describe the running of the anomalous dimension of the fermion bilinear, $\gamma$. Extending the perturbative result into the non-perturbative regime is of course a guess as to the strong coupling dynamics but we hope that that guess is a sensible choice as one approaches the point where $\gamma=1$ and the BF bound is violated.  In QCD-like theories where the coupling runs quickly at the BF bound scale the precise choice of running (e.g. one loop versus two loop) is a few percent effect but in more walking theories the ansatz is important. In the work here the running of the theories is fast and the $N_c$ dependence we report can be used as a measure of the errors on the choice of running. The key benefit of this modelling is that it allows the study of the explicitly dependence on the number of colours and flavours.  Since the model is based on probe branes it does not include backreaction on the geometry with the benefit that the model is quick to compute with. We note for comparison that a previous holographic model for Sp(2$N_C$) theories can be found in \cite{Imoto:2009bf} based on the Sakai Sugimoto model of QCD \cite{Sakai:2004cn}. That model includes the confinement scale via a compact extra dimension which then dictates the chiral symmetry breaking behaviour. Our model is based on the running of the anomalous dimension of the quark bilinear in the theory directly without the role of the extra dimension.

The model, which we have previously called Dynamic AdS/YM, was originally used for theories with mass degenerate fermions. Here one can study a single fermion flavour knowing that other states' masses are in degenerate multiplets with the flavour studied. We studied the spectrum of SU($N_c$) gauge theories with varying number of flavours \cite{Alho:2013dka, Erdmenger:2014fxa} seeking how the spectrum responded to the two loop IR walking fixed point behaviour. These results formed the basis of phenomenological studies of walking technicolour in \cite{Belyaev:2018jse}. More recently we studied the spectrum of theories with two different fermion representations \cite{Erdmenger:2020lvq,Erdmenger:2020flu} --- theories that can also underly composite Higgs models. Here we argued there could be gaps in scale between the bound states of the different representations \cite{Evans:2020ztq}.

Most recently \cite{Erdmenger:2023hkl} we have extended the model to non-abelian flavour symmetries following the spirit of the non-abelian DBI action \cite{Erdmenger:2007vj}. Here the full non-abelian flavour symmetry of the massless theory must be represented by a dual non-abelian gauge field in the bulk. Any mass terms, including the ones that break the flavour symmetry, are dynamical solutions of fields in the bulk and there is a bulk Higgs mechanism that must be carefully taken into account. In \cite{Erdmenger:2023hkl} we worked through the details of this construction including developing numerical methods to study cases where there is mixing between states as a result of the symmetry breaking. In this paper we will use this technology to study the many aspects of the Sp(2$N_c$) theory with mass flavour splitting.

Finally in another set of papers \cite{Evans:2016yas,Clemens:2017udk} we have studied including NJL interactions for the fermions of these models using Witten's multi-trace prescription \cite{Witten:2001ua}. One reinterprets massive solutions in the dual theory as one's with a condensate and a four-fermion operator that generates the mass dynamically. The critical coupling familiar from the NJL model is observed. In \cite{Clemens:2017luw} we used this methodology to study top condensation and top color assisted technicolour models.

Here, in this paper, we will bring all of this technology together to be applied to the Sp(2$N_c$) theory with four Weyl fermions. We will study the model with fermion masses (including the case of split masses for the left and right handed fields) compatible with leaving a light composite Higgs four-plet. We will include NJL interactions competing with the mass term to favour an alternative vacuum. We see a rotation from the composite Higgs vacuum to technicolour-like vacuum as observed field theoretically in \cite{Cacciapaglia:2014uja,Cacciapaglia:2020kgq}.  Here one can see that the intermediate regime between the composite Higgs and technicolour theories is very fine tuned in the coupling strength of the NJL interaction. The holographic model originates from large $N_c$ and does not naively include the U(1)$_A$ anomaly as we mentioned above. The anomaly has been included holographically in \cite{Barbon:2004dq} and could be included in future work. In practice this simply means our model will have one extra pseudo-Goldstone boson that in the true model will be more massive --- we will note this with our results below. 

At each stage we predict the bound state meson spectrum and decay constants. These computations show the power of the holographic model to include both the base chiral symmetry breaking gauge dynamics and that from NJL operators, but we also hope the spectrum predictions will help phenomenological searches for composite Higgs models. To date there has been  some lattice work on these models \cite{Arthur:2016dir,Arthur:2016ozw,Bennett:2019jzz, Bennett:2023rsl,Bennett:2023qwx,Kulkarni:2022bvh} although the lattice errors remain sizable. At this point the lattice results and our holographic predictions both share the same basic pattern of masses and couplings --- we hope that further more precise interplay between the lattice and holographic predictions will result in the future.

\section{The gauge theory}

We will consider the dynamics of an Sp($2N_c$) gauge theory with 2 Dirac fundamentals with the Lagrangian  \cite{Ryttov:2008xe, Cacciapaglia:2014uja,Cacciapaglia:2020kgq}
\begin{equation} 
{\cal L} = - {1\over 4} G^{\mu \nu}G_{\mu \nu} + i \bar{\psi}_i D \hspace{-0.25cm} \slash \psi_i - \bar{\psi}_i M_{ij} \psi_j. 
\end{equation}

\subsection{U(4) global symmetry}
Given the pseudo-real nature of the gauge theory we can write the fermion fields in terms of four two-component spinors and it is helpful to pick the naming convention
\begin{equation} \label{ewposition}
\psi_i = \left(   \begin{array}{c} U_L^C \\D_L^C\\ D_R \\U_R \end{array} \right) ,
\end{equation}
where we have written everything as right handed spinors by conjugating the left handed spinors. 

The theory has a U(4) global symmetry with generators
\begin{equation}
    \label{eq:U4_basis}
    T^{1-3}={1 \over 2 \sqrt{2}} \begin{pmatrix}
    \tau_i & 0 \\
    0 & \tau_i
\end{pmatrix}, \quad T^{4-6}={1 \over 2 \sqrt{2}}\begin{pmatrix}
    \tau_i & 0 \\
    0 & -\tau_i
\end{pmatrix},
\end{equation}
for $i=1,2,3$, $\tau_i$ are the Pauli matrices and
\small
\begin{equation} 
T^{7/8} = {1\over2\sqrt{2}}\left( \begin{array}{cccc} 0 & 0 & 0 & - i \\
0&0&\pm i&0 \\ 0 &\mp i & 0 & 0 \\  i & 0 &0 &0  \end{array} \right), \quad
T^{9/10} = {1\over2\sqrt{2}}\left( \begin{array}{cccc} 0 & 0 & - i & 0 \\
0&0&0&\mp i \\ i & 0 & 0 & 0 \\ 0 & \pm i &0 &0  \end{array} \right), 
\end{equation}
\begin{equation} 
T^{11/12} = {1\over2\sqrt{2}}\left( \begin{array}{cccc} 0 & 0 & 1 & 0 \\
0&0&0&\pm1 \\1 & 0 & 0 & 0 \\ 0 & \pm1 &0 &0  \end{array} \right), \quad
T^{13/14} = {1\over2\sqrt{2}}\left( \begin{array}{cccc} 0 & 0 & 0 & 1 \\
0&0&\pm 1&0 \\ 0 & \pm 1 & 0 & 0 \\ 1 & 0 &0 &0  \end{array} \right) , 
\end{equation}
\normalsize
\begin{equation} 
T^{15}={1\over 2\sqrt{2}}\begin{pmatrix}
    \mathbbm{1}_2 & 0 \\
    0 & -\mathbbm{1}_2
\end{pmatrix},\quad
T^{16}={-1\over 2\sqrt{2}}\begin{pmatrix}
    \mathbbm{1}_2 & 0 \\
    0 & \mathbbm{1}_2
\end{pmatrix}\,.
\end{equation}
Here if we were to consider embedding the theory in the standard model we might promote the left handed $U,D$ doublet to be in the fundamental representation of the gauged SU(2)$_L$ of the weak force. The 
corresponding generators are then  
$(T^1+ T^4)/\sqrt{2}$, $(T^2+T^5)/\sqrt{2}$ and
$(T^3+T^6)/\sqrt{2}$. We will always consider this gauging as a  neglectably weak perturbation on the strong gauge dynamics.

\subsection{Perturbative running}

Perturbatively the theory is asymptotically free for all $N_c$ and has beta function coefficients at one and two loops
\begin{equation}
    \label{eq:running}
    \begin{aligned}
        \mu{d\a\over d\mu}&=-b_0\a^2-b_1\a^3,\\
        b_0&={1\over 6\pi}\lrb{11C_2(G)-2\sum_R T(R)N_f(R)},\\
        b_1&={1\over 24\pi^2}\lrb{34C_2^2(G)-\sum_R\lrmb{10C_2(G)+6C_2(R)}T(R)N_f(R)},
    \end{aligned}
\end{equation}
where $R$ is the representation of the matter field. Here we have rewritten the expressions in terms of the number of Weyl fermions. The one loop relation for the anomalous dimension of the fermion bilinear is given by
\begin{equation} \label{gamma}
    \gamma = {3 C_2(R) \over 2 \pi} \alpha.
\end{equation}
In the above relations, $C_2$ is the quadratic Casimir, $G$ is the adjoint representation, $N_f$ is the number of flavours, $d(R)$ is the dimension of fermion's representation. The exact definitions for these factors of SU(N), SO(N) and Sp(2N) gauge theories are listed in \cite{Erdmenger:2020flu}, we list the relevant definitions in table \ref{tab:representations}.  
We set $\alpha(1)=0.65$ in our computations discussed below but then normalize all scales in terms of the physical $\rho$ mass so this choice is arbitrary and does not effect the outputs.
\begin{table}[t]
    \centering
    \begin{tabular}{|c|c|c|c|c|c|}
    \hline
         Gauge Group & $C_2(G)$ & $C_2(F)$             & $d(F)$ & $T(F)$ & $N_f=N_{\bar{f}}$ \\
         \hline
         $SU(N_c)$   & $N_c$    & ${N_c^2-1\over 2N_c}$& $N_c$  & $\half$ & 2\\
         \hline
         $Sp(2N)$    & $N$+1    & ${2N+1\over 4}$      & 2$N$   & $\half$ & 4\\
         \hline
    \end{tabular}
    \caption{Factors in eq.~\eqref{eq:running} for various gauge groups. The matter fields are in the fundamental representations, which is marked as $F$.}
   \label{tab:representations}
\end{table}

We note that using the anomalous
dimension at two-loop level does not lead to 
qualitative different results \cite{Erdmenger:2020lvq} in the model we present. Another check of the dependence in these theories on the choice of running is to use the $N_c$ dependence of the results as a measure of the effect of different ansatz for the running --- we will see the dependence is weak.

\subsection{Parametrizing fermion bilinear operators}

We will be interested in bound states of two fermions in the theory and also condensation of bilinear operators. It is helpful to parametrize the anti-symmetric bi-fermion operators as
\footnotesize
\begin{equation} \label{fluc1}
X_f= \left( \begin{array}{cccc}
0 & \sigma -Q_5 +i S -i \pi_5 & Q_2 - \pi_2 + i \pi_1 - i Q_1 & -Q_4 + \pi_4 +i Q_3 - i \pi_3 \\
 - \sigma+Q_5 +i \pi_5 - iS & 0 & Q_4 + \pi_4 +i Q_3 +i \pi_3 & Q_2 + \pi_2 + i Q_1 + i \pi_1 \\
 \pi_2 - Q_2 +i Q_1-i\pi_1 & -Q_4 - \pi_4 -i Q_3 - i \pi_3 & 0 & \sigma + Q_5 +i S + i \pi_5 \\
 Q_4-\pi_4 +i \pi_3 - i Q_3 & -Q_2-\pi_2 -i Q_1 -i \pi_1 & -\sigma - Q_5 -i S - i \pi_5 & 0 
\end{array} \right).
\end{equation}
\normalsize
$X$ transforms under the U(4) flavour symmetry $\Omega$  as
\begin{align} \label{breakcondition}
X_{ab} \to X_{cd}' = \Omega_{ca} \Omega_{db} X_{ab}
\Leftrightarrow X \to X' = \Omega X \Omega^T \,.
\end{align}

\subsection{Fermion condensation}
The expectation is that when these gauge theories reach strong coupling, as in QCD, bi-fermion operator condensation will occur. We assume the theories do not break their own colour groups so the bilinear must be a colour singlet which is always anti-symmetric in colour indices for these theories. We likewise assume the theory does not break Lorentz invariance and so the angular momentum wave function must be antisymmetric. The conclusion then is that by Fermi-Dirac statistics the condensate that forms must be anti-symmetric in flavour space also. For example the 4$\times$4 matrix $X$ might take the form (here the field $\sigma$ has acquired a vacuum expectation value (vev)) 
\small \begin{equation} \label{1stvacnow}
    X=\begin{pmatrix}
        0  & L_0 & 0 & 0\\
       -L_0 & 0  & 0 & 0\\
        0 & 0 & 0 & L_0 \\
        0 & 0 & -L_0 & 0\\
    \end{pmatrix}.
\end{equation}
\normalsize
This vev is invariant under an Sp(4) sub-group of U(4). This can be seen explicitly by looking for invariance under eq.~\eqref{breakcondition}. The broken generators are $T^i$ for $i=8,10,11,14,15,16$. U(4) has 16 generators whilst Sp(4) has 10. There are 6 broken generators and there will be 6 Goldstone bosons --- the scalars $\pi_i$ for $i=1,...,5$ and S. The $S$ state would be made massive  by the anomaly were it included.

Equally one can consider an equivalent vacuum where $Q_4$ acquires a vev
\small
\begin{equation} \label{Xchnow}
    X_{Q_0}=\begin{pmatrix}
        0  & 0 & 0 & -Q_0\\
       0 & 0  & Q_0 & 0\\
        0 & -Q_0 & 0 & 0 \\
        Q_0 & 0 & 0 & 0\\
    \end{pmatrix}.
\end{equation}
\normalsize
 $X_{Q_0}$ can be transformed to the form in eq.~\eqref{1stvacnow} by $X = U X_{Q_0} U^T$ with 
\small
\begin{equation} \label{Xtc}
    U = {1 \over \sqrt{2}}\begin{pmatrix} 
    0 & -1 & ~0 & ~1 \\
    1 & ~0 & -1 & ~0 \\
    0 & ~1 & ~0 & ~1 \\
    1& ~0 & ~1 & ~0
    \end{pmatrix} \,.
\end{equation}
\normalsize
Again U(4) flavour is broken to Sp(4) and there are 6 Goldstone bosons (now $Q_1$--$Q_5$ and $S$).

\subsection{Mass terms}

Fermion mass terms can be introduced in the same patterns as the vevs discussed already and they will tend to align the vacuum to the mass pattern. Thus for example the two condensate patterns above would be favoured respectively by the mass matrices

\small
\begin{equation} \label{masses}
    M=\begin{pmatrix}
        0  & m_1 & 0 & 0\\
       -m_1 & 0  & 0 & 0\\
        0 & 0 & 0 & m_2 \\
        0 & 0 & -m_2 & 0\\
    \end{pmatrix}, \quad
     M=\begin{pmatrix}
        0  & 0 & 0 & -m_1\\
       0 & 0  & m_2 & 0\\
        0 & -m_2 & 0 & 0 \\
        m_1 & 0 & 0 & 0\\
    \end{pmatrix}.
\end{equation}
\normalsize
In each case with $m_1=m_2$ the flavour symmetry breaking pattern U(4)$\rightarrow$Sp(4) is explicit and the Goldstone modes will become pseudo-Goldstones with small mass squared proportional to the fermion mass. In the mass split case $m_1 \neq m_2$ the global symmetry is explicitly broken to SU(2)$_L \times$ SU(2)$_R$.

Below we will find it useful that the mass or vev matrix
\small
\begin{equation} \label{Xch}
    X=\begin{pmatrix}
        0  & L_0 & 0 & -Q_0\\
       -L_0 & 0  & Q_0 & 0\\
        0 & -Q_0 & 0 & L_0 \\
        Q_0 & 0 & -L_0 & 0\\
    \end{pmatrix},
\end{equation}
\normalsize
can be placed, by a transformation $X \rightarrow U X U^T$, to the forms 
\small
\begin{equation} \label{trans1}
    X \rightarrow  \begin{pmatrix}
        0  & L_0+Q_0 & 0 & 0\\
       -L_0-Q_0 & 0  & 0 & 0\\
        0 & 0 & 0 & L_0 -Q_0\\
        0 & 0 & -L_0+Q_0 & 0\\
    \end{pmatrix}
, \quad{\rm using}\quad
    U = {1\over \sqrt{2}}\begin{pmatrix} 
    ~0 & -1 & ~0 & ~1 \\
    ~1 & 0 & -1 & ~0 \\
    ~0 & ~1 & ~0 & ~1 \\
    -1& 0 & -1 & ~0
    \end{pmatrix},
\end{equation}
\normalsize

\small
\begin{equation} \label{trans2}
    X \rightarrow  \begin{pmatrix}
        0  & L_0+Q_0 & 0 & 0\\
       -L_0-Q_0 & 0  & 0 & 0\\
        0 & 0 & 0 & -L_0 +Q_0\\
        0 & 0 & L_0-Q_0 & 0\\
    \end{pmatrix}
, \quad {\rm using} \quad
    U = {1 \over \sqrt{2}}\begin{pmatrix} 
    0 & -1 & ~0 & ~1 \\
    1 & ~0 & -1 & ~0 \\
    0 & ~1 & ~0 & ~1 \\
    1& ~0 & ~1 & ~0
    \end{pmatrix} \,.
\end{equation}
\normalsize

\subsection{Composite Higgs to Technicolour}

The vacua discussed in eq.~\eqref{1stvacnow} and eq.~\eqref{Xchnow} are formally equivalent in the pure strongly coupled massless system. If though we were to gauge SU(2)$_L$ in the basis eq.~\eqref{ewposition} then the first case eq.~\eqref{1stvacnow} leaves the weak force unbroken whilst the second eq.~\eqref{Xchnow} breaks it (in a QCD-like or technicolour-like pattern). The effective potential of the theory would prefer not to break gauge fields and the first vev would be preferred \cite{Peskin:1980gc}. 

This case, which can also be favoured by the electroweak preserving masses shown on the left in eq.~\eqref{masses}, serves as the basis of some composite Higgs models. The Goldstone $\pi_1-\pi_4$ fields are doublets of SU(2)$_L$ and SU(2)$_R$ and in these models are interpreted as the Higgs (with top loops generating an effective potential that breaks electroweak symmetry \cite{Arkani-Hamed:2002ikv,Golterman:2015zwa}). 

Alternatively it is possible to favour the technicolour like vev of eq.~\eqref{Xchnow} by including NJL four-fermion interactions. The Lagrangian terms
\begin{equation} \label{NJLops}
    {\cal L} = {g_s^2 \over \Lambda_{UV}^2 } ( \bar{\Psi}_L U_R \bar{U}_R  \Psi_L + \bar{\Psi}_L D_R \bar{D}_R  \Psi_L),
\end{equation}
where $\Lambda$ is the UV cut-off and $g_s$ the dimensionless NJL coupling for this scalar operator squared, preserve electroweak symmetry but support the condensates in eq.~\eqref{Xchnow}. These operators would naturally be present for example if the gauge group were SU(2)=Sp(2) and embedded at higher scales into an SU(N) theory where the pseudo-reality condition is not present. 

In  \cite{Ryttov:2008xe, Cacciapaglia:2014uja,Cacciapaglia:2020kgq} the transition between these limits was explored field theoretically. In this paper we will study the strong dynamics using holography including the NJL terms and masses. We will be able to make predictions for the masses of mesonic bound states of the theory as a function of the fermion masses and NJL coupling. 

\section{Holographic model}

To holographically describe the gauge theory we will use our Dynamical AdS/YM model. It was developed and used in the abelian case in \cite{Alho:2013dka,Erdmenger:2020flu} and extended to the non-abelian case in \cite{Erdmenger:2023hkl}. The model is based on the canonical example of including flavour in AdS/CFT, the D3/probe D7 system \cite{Karch:2002sh,Kruczenski:2003be, Erdmenger:2007cm} and its non-abelian version \cite{Erdmenger:2007vj}. In that top down system chiral symmetry breaking is induced by elements of the gravitational background that generate an effective IR BF bound violating mass for the scalar in the DBI action describing the D7-brane position (dual to the chiral condensate) \cite{Alvares:2012kr}. The philosophy is to simply import the kinetic terms of the D7-brane DBI action but by hand input the running of the fermion bilinear's anomalous dimension through the mass squared of the scalar field (using the standard dictionnary relation $\Delta (\Delta-4)=M^2$).

\subsection{The $X$ field}

Here we will include the bilinear operator vevs  and fluctuations  through a scalar field $X$ in the bulk with the flavour structure we have already seen in the field theory eq.~\eqref{fluc1}. These fields will live  in the AdS$_5$ space
\begin{equation}
    ds^2 = (\rho^2 + X^\dagger X) dx_{3+1}^2 +  (\rho^2 + X^\dagger X)^{-1}d\rho^2.
\end{equation}
$\rho$ is the holographic radial direction of AdS dual to RG scale. Note here the metric components are flavour matrices. The X fields enter the metric feeding knowledge of the mass gap of the theory to the fluctuation modes (as happens in the D3/probe D7 system). 

Let us begin by discussing the $X$ field theory before we add the U(4) gauge fields.  The action for the $X$ fields is given by
\begin{equation}
    \label{eq:proposed_lag} 
    \begin{aligned}
    {\cal L}  & = { 1 \over 2} \rho^3 \Tr \lrb{\partial_\rho X^\dagger \partial_ \rho X} + {1 \over 2} \rho^3 \Tr \lrb{\rho^2+X^\dagger X }^{-2} \lrb{\partial_x X^\dagger \partial_x X} \\&  + {1 \over 2}\rho ~A\lrmb{\Tr \lrb{\rho^2 + X^\dagger X} }\Tr X^\dagger X + {B  \over  \rho} \lrb{4 \Tr  X^\dagger X  X^\dagger X - \lrb{\Tr  X^\dagger X}^2} \,.\end{aligned} 
\end{equation}

\noindent The terms in the first line here are simply the non-abelian DBI kinetic terms for the field $X$ with appropriate metric factors for the $\rho$ and spatial $x^\mu$ derivative terms. The first term in the second line is a $\rho$ dependent (RG scale dependent) mass term for $X$. $A$ is indicated as a function of $\Tr (\rho^2 + X^\dagger X)$. At $X=0$ we use this function  to input the gauge theory's running anomalous dimension.   As written, all fields (generically $f$) in X will have an equation of motion (neglecting $B$) that looks like
\begin{equation}
    \label{eq:general_f_eom}
    \partial_\rho (\rho^3 \partial_\rho f) - \rho A f - {1 \over 2} {\partial A \over \partial f}  f^2 = 0.
\end{equation}
To impose the equation of motion we want we set
\begin{equation}
    \label{eq:id_dm}
    A  + {1 \over 2}{\partial A \over \partial f}  f \equiv \Delta m^2 \lrmb{\rho^2+f^2} \,.
\end{equation}
Here $\Delta m^2$ is an effective correction to the mass that depends on $r^2=\rho^2+f^2$. Note that since it is easier to solve the running of $\alpha$ against $\log \mu$ it is sometimes helpful to think of $\Delta M^2$ being a function of $\log \mu =  \log r = \log ( \sqrt{\rho^2 + f^2}$).
Note $f$ enters this function to allow a stable non-BF bound violating solution to form at low $\rho$ if $f$ develops a vev.  A more careful analysis of this identification in the full model is in appendix \ref{app:dm_non_deg}.

Now we can explicitly input the dynamics of a particular gauge field. We use the standard scalar AdS/CFT dictionary $M^2=\Delta(\Delta-4)$ to relate $\Delta m^2$
to the gauge theory running. Note that to use the naive mass dimension relation one must write the field $f$ as $\rho \phi$ and then one sees $\phi$ has a mass squared term of $-3$ before we introduce $\Delta m^2$.
Now we can explicitly input the dynamics of a particular gauge field. Using $M^2=\Delta(\Delta-4)$ for small $\Delta$, the correct relation one must impose is \cite{Alho:2013dka}
\begin{equation}
    \Delta m^2 = - 2 \gamma,
\end{equation}
where we take $\gamma$ from eq.~\eqref{gamma} and relate the AdS radial direction directly to the RG scale so $\rho^2+X^\dagger X = \mu^2$. Of course, using the perturbative runnings into the non-perturbative regime is an assumption but we hope that the two loop running of $\alpha$ captures the important features of the running (including IR walking behaviour) as one approaches the point where the BF bound is violated at $\gamma=1$ and that the features of the pole are not crucial to the spectrum. Again we note that the $N_c$ dependence of the results provides some gauge of how the choice of running changes the results.

So far our action is only a function of $(\Tr  X^\dagger X)^n$. Were we to leave it in this form then there would be an accidental symmetry that would make the 12 elements of $X$ a 12-plet of SO(12). To enact that it is truely a bi-fundamental of flavour we must also include terms with trace form $\Tr [(X^\dagger X )^n]$.
The quartic $B$ term is constructed as the simplest such term that incorporates the additional symmetry breaking in the 12-plet of $X$. The quintuplet fields $Q_i$ and $\sigma$  are present in every term  and that breaks the symmetry between those fields and the $\pi_i$ and $S$ fields. 

A downside of having to include the quartic $B$ term is that we have added an extra constant that will enter the $Q_i,\sigma$ masses. We're no longer predicting their mass splitting from $\pi_i$ and $S$ and equally won't be able to fully track their $N_c$ dependence since we don't know how $B$ depends on $N_c$.

\subsection{Condensates, mass terms and NJL interactions}

The $X$ field allows us to include mass terms for the 12 operators, their matrix forms described in the usual AdS/CFT dictionnary fashion. The field has the UV asymptotic behaviour 
\begin{equation} \label{mc}
    X_{ij} \sim {\cal J} + {{\cal O} \over \rho^2}=  m_{ij} + \bar{\psi}_i \psi_j/ \rho^2. 
\end{equation} 
Mass terms (the source ${\cal J}$) are therefore included by fixing the UV boundary terms in the vacuum solution describing the operator vevs, ${\cal O}$. We find it useful to distinguish the vev of $X$ from the fluctuations so will write
\begin{equation}
    \label{eq:X}
    X=X_0+X_f,
\end{equation}
where $X_0$ will be a matrix describing the vacuum values of $\sigma=L_0(\rho)$ and $Q_4=Q(\rho)$ --- we will explore vacuums of the form seen in eq.~\eqref{1stvacnow}, eq.~\eqref{Xchnow} and eq.~\eqref{Xch} plus masses generically as on the left in eq.~\eqref{masses}. 
$X_f$ is the scalar fluctuations parametrised for example as in eq.~\eqref{fluc1}.

In addition though Witten's multi-trace prescription \cite{Witten:2001ua} allows us to include NJL four-fermion operators
\begin{equation}
    {\cal L} = {g_s^2 \over \Lambda^2} \bar{\psi}_i \psi_j \bar{\psi}_j \psi_i,
\end{equation} 
where $\Lambda$ is the UV cut-off and $g_s$ the dimensionless NJL coupling for this scalar operator squared. This was explored in detail in the D3/D7 system in \cite{Evans:2016yas,Clemens:2017udk} and can be imported directly here. We allow the same massive solutions already described in the theory but reinterpret them. We now assume that any mass present is due to the four-fermion operator and the vev so
\begin{equation} \label{NJLbc}
    {g_s^2 \over \Lambda^2}  = {{\cal J} \over {\cal O} },
\end{equation}
where ${\cal J}$ is the source (the mass) and ${\cal O}$ the operator vev in eq.~\eqref{mc}. 

\subsection{The U(4) gauge field}

We will also now include a bulk gauge field for the U(4) flavour symmetry describing the $\bar{\psi} \gamma^\mu \psi$ operators and the associated vector mesons. The U(4) gauge field, $A^b$ with $b=1,...,16$ is associated with the generators  in eq.~\eqref{eq:U4_basis}. In addition to its kinetic term, this gauge field will couple to X via the covariant derivative
\begin{equation}
    \label{eq:covariant_d}
    D^M X=\pd^M+i A^{bM}(T^{b}X+XT^{bT}).
\end{equation}
The full action in the bulk is now
\begin{equation}
    \label{eq:action}
    \begin{aligned}
     S=&\int d^4x d\rho 
\rho^3 \STr\biggl\{{1\over r^2}(D^M X)^\dagger D_M X+{1\over\rho^2} A\lrmb{Tr (\rho^2 + X^\dagger X) }Tr X^\dagger X\\
&+ {B  \over \rho^4} (4 \Tr  X^\dagger X  X^\dagger X - (\Tr  X^\dagger X)^2)+{2\over g_5^2}\lrb{F_{A,ab}F_A^{ab}}\biggr\},
    \end{aligned}
\end{equation}
where $M=1,...,5$ is the spacetime index.  $\STr$ is the symmetrised trace, as defined in \cite{Tseytlin_1997}.  The 5-dimensional coupling $g_5^2$ is 
\begin{equation}
    \label{eq;g5sq}
    g_5^2={24\pi^2\over d(R) N_f(R)},
\end{equation}
which, as usual in AdS/QCD models, is fixed to give the correct UV vector vector source correlator --- see \cite{Erlich:2005qh, Alho:2013dka}.

\subsection{Fluctuations}

To find the mass spectrum of the theory we will seek fluctuations about the vacuum configuration. We expand the action to quadratic order and solve the linear equations of motion that result for generic fluctuation fields $f_i$.
We look at fluctuations with space-time dependence $f_i(\rho) e^{ik.x}$, $k^2=-M^2_f$, the natural UV boundary conditions are that $X_f \rightarrow 0$ in the UV so that the fluctuations are only of the operator. 

Thus to compute the masses numerically, we adopt the following boundary conditions using the shooting method for the decoupled equations of a general field denote with $f$
\begin{equation}
    \label{eq:deg_bc_single}
    f(\rho_{IR})=1,\quad\pdr f(\rho_{IR})=0, \quad f(\rho_{UV})=0,
\end{equation}
where the shooting parameter is the bound states' masses $M_f$. Where we encounter coupled equations of two fields, denoted as $f_1$ and $f_2$, the boundary conditions are \cite{Erdmenger:2023hkl}
\begin{equation}
    \label{eq:deg_bc_coupled}
    \begin{aligned}
        &f_1(\rho_{IR})=1,\quad\pdr f_1(\rho_{IR})=0,\quad f_1(\rho_{UV})=0,\\
        &f_2(\rho_{IR})=b,\quad\pdr f_2(\rho_{IR})=0,\quad f_2(\rho_{UV})=0,
    \end{aligned}
\end{equation}
where in this case the common mass $M_{f}$ and the boundary value $b$ at the IR are the shooting parameters used to achieve the UV boundary conditions. These shooting methods are implemented using the \verb  |NDSolve| function from \verb|MATHEMATICA|. 

A final caveat is that one must be careful when studying the fluctuations when NJL terms are present. In that case the fluctuations must see the appropriate UV boundary condition satisfying the same NJL coupling relation between operator and source as the background eq.~\eqref{NJLbc}. We will see examples of both mass and NJL cases (and a combination) below.

\subsection{Decay constants}

The decay constants for the mesons are obtained by calculating two-point correlation functions. This requires coupling the bound state to the corresponding external source. The external sources are fluctuation solutions with a non-normalisable term in the UV. We write the fluctuations  as $L_0=K_S(\rho)e^{-iq\cdot x}$ for the scalar and $V^\mu=\epsilon^\mu K_V(\rho)e^{-iq\cdot x}$ for the vector. Their equations in the UV can be summarised as \cite{Alho:2013dka,Erdmenger:2020flu}
\begin{equation}
    \label{eq:fluc_uv_for_fv}
    \pdbr{K}-{q^2\over \rho}K=0.
\end{equation}
The solutions in the UV is 
\begin{equation}
    \label{eq:f_sol_K}
    K_i=N_i\lrb{1+{q^2\over4\rho^2}\ln\lrb{q^2\over\rho^2}}, \quad i=S,V,A.
\end{equation}
$N_i$ are the normalization constants that are determined by matching to the UV gauge theory \cite{Erlich:2005qh,Alho:2013dka}
\begin{equation}
    \label{eq:f_Ni}
    N_S^2={d(R) N_f(R)\over 48\pi^2},\quad N_V^2=N_A^2={g_5^2d(R)N_f(R)\over 48\pi^2}.
\end{equation}

Substituting the solutions of the fluctuations back into the action and integrating by parts, the overlap between the bound state fluctuation and the external source is 
\begin{equation}
    \label{eq:f_S}
    F_i^2=\int d\rho {1\over g_5^2}\pdr(-\rho^3\pdr f)K_i(q^2=0),\quad f,i=S,V,A.
\end{equation}
The pion decay constants are calculated from the overlap of two broken vector currents, i.e.
\begin{equation}
    \label{eq:f_pi}
    f_\pi^2=\int d\rho {1\over g_5^2}\pdr(-\rho^3\pdr K_A(q^2=0))K_A(q^2=0).
\end{equation}

\section{The mass degenerate theory}\label{sec:degenerate_case}

We will begin by describing the basic Sp(4) gauge theory with a common mass term linking each of two pairs of the fermion fields, including the massless limit. That is we will write the vacuum expectation value of the holographic field $X$ as 
\small
\begin{equation}
    \label{eq:vac_deg}
    L=\begin{pmatrix}
        0  & L_0(\rho) & 0 & 0\\
       -L_0(\rho) & 0  & 0 & 0\\
        0 & 0 & 0 & L_0(\rho) \\
        0 & 0 & -L_0(\rho) & 0\\
    \end{pmatrix} \,.
\end{equation}
\normalsize
We are thus choosing the basis in eq.~\eqref{1stvacnow} with the vacuum value of the holographic field $\sigma=L_0(\rho)$ --- the source value in the UV corresponds to the mass (in the case $m_1=m_2\equiv m$ in the left hand mass matrix of eq.~\eqref{masses}) and the UV operator corresponds to the vev of $L_0$ in eq.~\eqref{1stvacnow}.
This vev will break the U(4) global symmetry to Sp(4).

The scalar fluctuations, parameterised as in eq.~\eqref{fluc1}, split into: the Higgs-like mode $\sigma$ (a singlet of Sp(4)); the Goldstone modes   $\pi_{1,...,5}$ (a 5-plet of Sp(4)) and  $S$ (a singlet); and the modes made massive by the $B$-term  in eq.~\eqref{eq:action} $Q_{1,\dots, 5}$ (a 5-plet). 

The broken generators of U(4) are $T^i$ for $i=8,10,11,14,15,16$ in eq.~\eqref{eq:U4_basis} and the six associated gauge fields in the bulk will experience a Higgs mechanism. The vector fields split into a 10-plet adjoint of Sp(4), a 5-plet and a singlet.

\subsection{The vacuum}

\begin{figure}
    \centering
    \includegraphics{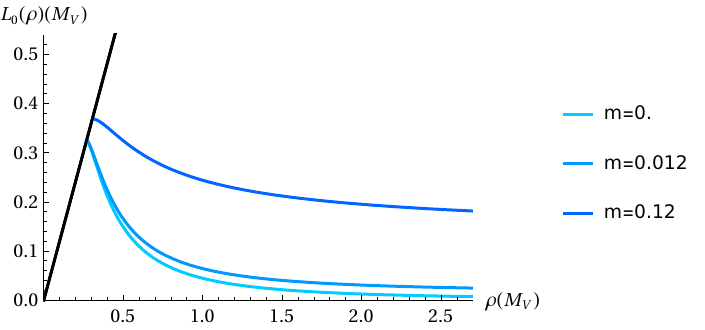}
    \caption{Vacuum solutions for $L_0$ against RG scale $\rho$ for different UV fermion masses in the SU(2) theory. The plot is in units where $M_V=1$ when $m=0$. The embeddings corresponds to $m$=0, 0.012 and 0.12 respectively.}
    \label{fig:deg_L0_vs_mq}   
\end{figure}

The equation of motion for the background is given by 
\begin{equation}
    \label{eq:deg_vac_eom}
     \partial_\rho ( \rho^3 \partial_\rho L_0)-\rho \Delta m^2 L_0 = 0, 
\end{equation}
where $\dmq$  is a function of $\log(\rho^2 + L_0^2)$.   We solve this equation using the \verb  |NDSolve| function from \verb|MATHEMATICA| with an IR boundary condition 
\begin{equation}
    \label{eq:deg_vac_bc}
    L_0(\rho_{IR})=\rho,\,\pdr L_0(\rho_{IR})=0,
\end{equation}
corresponding to the IR point where the fermions go on mass shell. By varying $\rho_{IR}$ one can achieve different UV values of $L_0$ corresponding to different fermion mass choices. We show some example solutions with different UV choices of the fermion mass in figure~\ref{fig:deg_L0_vs_mq}.

\subsection{Fluctuations --- the spectrum}

The $\s$ fluctuation around $L_0(\rho)$ is given by the solution of
\begin{equation} \label{eq: eqn of motion_scalar} 
\partial_{\rho} \lrb{\rho^3 \partial_{\rho} \sigma(\rho)} - \rho \Delta m^2 \sigma(\rho) - \rho L_{0}(\rho) \sigma(\rho) \frac{\partial \Delta m^2}{\partial L} \biggr\vert_{L_{0}} + M^2 \frac{\rho^3}{r^{4}} \sigma(\rho) = 0, 
\end{equation}
where $r^2=(\rho^2 + L_0^2)$.

The $Q_i$ fields satisfy
\begin{equation}
    \label{eq:deg_Q_eoms}
    \partial_\rho ( \rho^3 \partial_\rho Q_i) -\rho \Delta m^2 Q_i - {16B  \over \rho} L_0^2 Q_i + M^2 \frac{\rho^3}{r^{4}} Q_i= 0.
\end{equation}

There are ten unhiggsed  vector mesons that  satisfy the equation of motion
\begin{align} \label{eq: eqn of motion_vector}
\partial_{\rho} (\rho^3 \partial_{\rho}V_i(\rho)) + M^2_{V_i} \frac{\rho^3}{r^{4}} V_i(\rho) = 0,\quad i= 1-7, 9, 12,13.
\end{align}

There are 6 higgsed ``axial"-vector mesons  described by the $\vec{x},t$ components of $A^N$ by the equations
\begin{equation}  \label{eq: eqn of motion_axial}
\partial_{\rho} \lrb{\rho^3 \partial_{\rho} A_i(\rho)} - g_5^2 \frac{\rho^3 L^2_{0}}{r^4} A_i(\rho) + \frac{\rho^3 M^2_{A_i} }{r^{4}} A_i(\rho) = 0,\quad i=8,10,11,14-16 . 
\end{equation}

To compute the $\pi$ and $S$ masses we should work in the $A_\rho =0$ gauge and write $A_\mu = A_{i,\mu\perp} + \partial_\mu \phi_i$, for $i=8,10,11,14-16$. The $\pi$ and $S$ fields are degenerate and they mix with $\phi$ to describe the (pseudo) Goldstone bosons, we have 
\begin{equation} \label{eq: pion_full}
\begin{split}
\partial_{\rho} \lrb{\rho^3 \partial_{\rho} \phi_j } + g_5^2 \frac{ L_0\rho^3 }{r^4} \lrb{\sqrt{2} \pi_i - L_0 \phi_j} =0& , \\
\partial_\rho \lrb{\rho^3 \partial_\rho \pi_i} -\rho \Delta m^2 \pi_i  + M^2 \frac{\rho^3 }{r^{4}} \lrb{\pi_i-{L_0\over\sqrt{2}}\phi_j}= 0&,\\
i=1-5,\,j=14,8,11,10,15&.
\end{split}
\end{equation}
Here the ordering of the $j$ values is given by the mixing of $\pi_i$ with $\phi_j$, e.g.
$\pi_1$ mixes with $\phi_{14}$ and so forth.
The $S$ field mixes with $\phi_{16}$ in the same way. The Goldstone nature of these fields can be seen explicitly. There is a $M^2=0$ solution to eq.~\eqref{eq: pion_full}  where $\pi=L_0$ and $\phi=\sqrt{2}$. In the theory where we enforce that the masses of the fields are zero, $L_0$ asymptotes to zero in the UV. This solution for the Goldstone is then a physical solution in the gauge theory (just an operator fluctuation).

If though $L_0$ asymptotes to a non-zero UV value we have included a fermion mass, and now the Goldstone solution we have identified is not a physical solution in the gauge theory because it corresponds to a space-time dependent fluctuation of the mass. Nevertheless that Goldstone mode is present in the bulk and higgses the bulk gauge fields to enact the symmetry breaking generating the vector  mass term. The physical pseudo-Goldstone mass in the gauge theory is found by solving eq.~\eqref{eq: pion_full} with the requirement $\pi_i$, $L_0\phi_i \rightarrow 0$ in the UV --- a non zero mass results. 

Finally though we can include a mass in the UV solution of $L_0$ but attribute it to an NJL term. The massless solution then returns to being a physical solution since using $L_0$ for the fluctuation then shares the NJL coupling of the background. The state is again a Goldstone. We will not explicitly explore the NJL version of the theory in this section though we will return to it later.

We solved the equations of motion of all the fluctuations using the boundary conditions eq.~\eqref{eq:deg_bc_single} and eq.~\eqref{eq:deg_bc_coupled} when $m \sim 0$, and $B=1$ for various choices of $N_c$. The results are listed in table~\ref{tab:deg_mass_tab}. At each value of $N_c$ we have rescaled the vector bound state's mass $M_{V_0}$ at $m=0$ to one and rescaled all the other masses in this unit. 

The expected six massless Goldstones ($\pi$ and $S$) are present. The remaining bound states' masses are reasonably compatible with the data from a lattice simulation (although the errors on some of the simulations are large). The masses are fairly stable with changing $N_c$ although the axial vector meson masses fall somewhat with $N_c$ --- this drop is not seen in the current lattice data where the $Sp(4)$ data looks more like the holographic result at $SU(2)$. On the other hand our $\rho-A$ mass splitting is compatible within the lattice errors and we need to wait on improved lattice data to truly compare.  

\begin{figure}
    \begin{subfigure}[b]{0.45\textwidth}
        \centering
        \includegraphics[width=\linewidth]{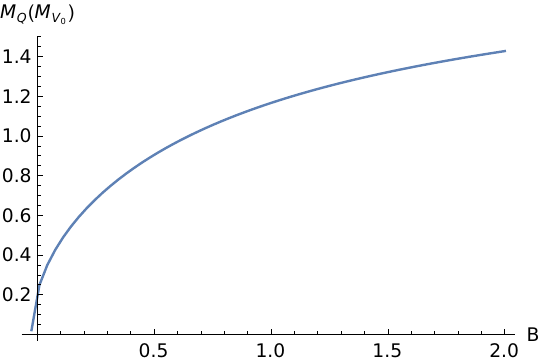}
        \caption{The Q mass increases with B.}\label{fig:mQ_vs_B_a}
    \end{subfigure}\hfill
    \begin{subfigure}[b]{0.5\textwidth}
        \centering
        \includegraphics[width=\linewidth]{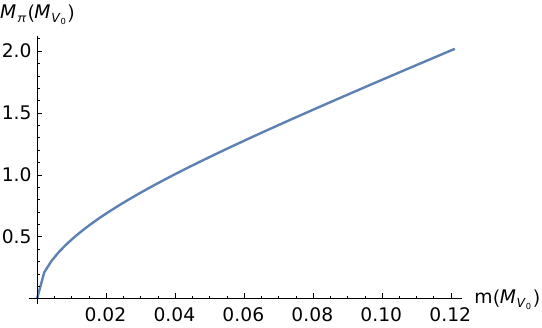}
        \caption{The degenerate $\pi$ states.}\label{fig:mQ_vs_B_b}
    \end{subfigure}\hfill
    \caption{ \label{fig:mQ_vs_B} Results for the SU(2) gauge theory. On the left: the Q state mass varies with the parameter $B$, $m=0.006 M_{V_0}$. Below $B\sim-0.03$ the mass becomes tachyonic, which is not favoured by the current discussion. On the right: The $\pi$ mass as a function of the UV fermion mass. As $m\rightarrow 0$ the $\pi$ mass vanishes, which shows the behaviour of a Goldstone boson.
    }
\end{figure}

The mass of the $Q$ fields are particularly dependent on our choice of $B$. If $B=0$ then the model develops an enhanced symmetry and the $Q$s become Goldstones degenerate with the $\pi$s. This is not expected in the theory that does not have this accidental symmetry. In figure~\ref{fig:mQ_vs_B_a} we plot the $Q$ mass against $B$ in the SU(2) theory and show that the mass has stabilized around $B=1$ to around the scale of the other hadrons although we can not predict its precise value. 
\begin{table}
    \centering
    \begin{tabular}{|c|c|c|c|c|c|c|c|}
    \hline
        Observables & $SU(2)$  &Lattice $SU(2)$& $Sp(4)$ & Lattice $Sp(4)$ &  $Sp(6)$& $Sp(8)$ &$Sp(10)$ \\
        \hline
         $m_V$ (10)   & $1^*$ & 1.00(3) & $1^*$  & 1.00(33) & $1^*$& $1^*$& $1^*$\\
         $m_A$  (6)  & 1.66  & 1.11(46)& 1.26& 1.61(17) & 1.18 & 1.14  &1.12\\
         $m_\s$ (1)  & 1.26  & 1.5(1.1)& 1.20  & &   1.22 & 1.23 & 1.23\\
         $m_Q$ (5)   & 1.13  &         & 1.13 &   & 1.13  & 1.13 & 1.13\\
         $m_{\pi,S}$ (6)& 0.02  &      & 0.01 & & 0.01  & 0.01 & 0.01 \\
         $F_V$       &   0.38   &      & 0.53& 0.52(10) &0.59  & 0.64 & 0.67\\
         $F_A$       &   0.48   &      & 0.54 & 0.673(92)&  0.59  & 0.63 & 0.66\\
         $f_\pi$     &   0.06   &     & 0.10 & 0.122 (99) & 0.12  & 0.12 & 0.13\\
         \hline
    \end{tabular}  
    \caption{Bound states' masses and decay constants at $m\sim0$ for gauge groups SU(2) to Sp(10) respectively. The 4 Weyl fermions are in the fundamental representations of each gauge group. We normalized the vector mass to 1 (this is noted by the asterisk). In computing we set $B=1$ as an example.  The lattice result for the SU(2) gauge theory are taken from \cite{Arthur:2016dir,Arthur:2016ozw}. The lattice results for the Sp(4) theory are from \cite{Bennett:2019jzz} --- the scalar sector has not yet been studied at low mass \cite{Bennett:2023rsl}. A further lattice study at large quark mass is in \cite{Kulkarni:2022bvh}.}
    \label{tab:deg_mass_tab}
\end{table}

\begin{figure}
    \centering
    \includegraphics{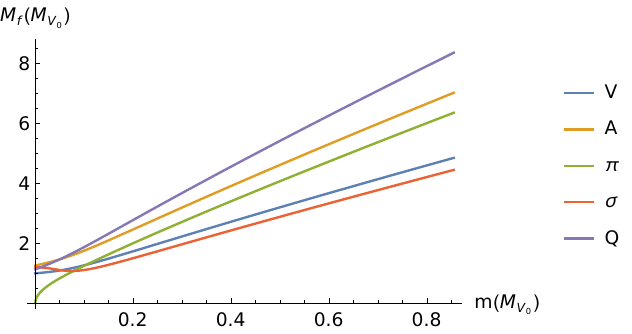}
    \caption{The masses of the bound state $f$ against the UV fermion mass in the unit of $M_{V_0}$. The gauge group is Sp(4) and $B=1$. }
    \label{fig:deg_all_flucs_vs_mq}
\end{figure}

We can also plot the masses and decay constants as functions of the UV fermion mass for $B=1$. This is shown in figure~\ref{fig:deg_all_flucs_vs_mq} for the Sp(4) gauge group case. We show a close up of the Goldstone masses against fermion mass, showing the expected $M_\pi \sim \sqrt{m}$  behaviour at small $m$ in figure~\ref{fig:mQ_vs_B_b}. 
We note for completeness, that 
the qualitative behaviour is same in
case of Sp(4).
Otherwise the bound state masses in figure \ref{fig:deg_all_flucs_vs_mq} broadly just increase with $m$ --- the only notable difference is the $\sigma$ mass that slightly decreases at small $m$ before

\begin{figure}
    \centering
    \begin{subfigure}[b]{0.5\textwidth}
        \centering
        \includegraphics[width=\linewidth]{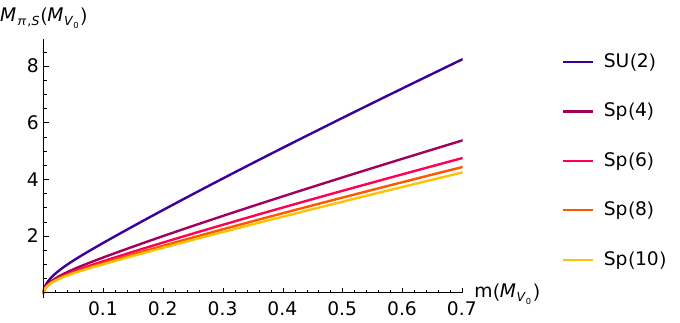}
        \caption{$M_{\pi,S}$}
    \end{subfigure}\hfill
    \begin{subfigure}{0.42\textwidth}
        \centering
        \includegraphics[width=\textwidth]{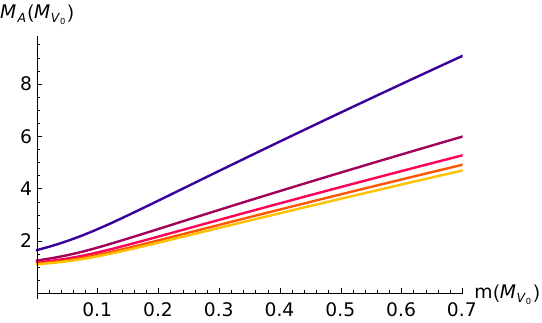}
        \caption{$M_A$}
    \end{subfigure}\hfill
     \begin{subfigure}{0.42\textwidth}
        \centering
        \includegraphics[width=\textwidth]{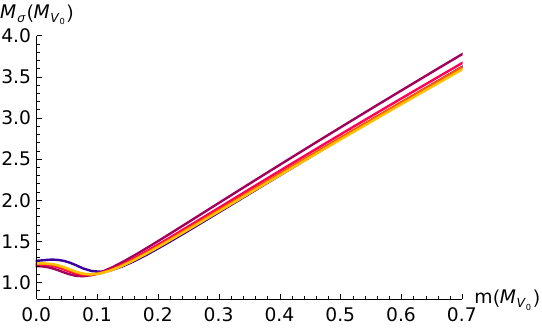}
        \caption{$M_\s$}
    \end{subfigure}\hfill
    \begin{subfigure}{0.42\textwidth}
        \centering
        \includegraphics[width=\textwidth]{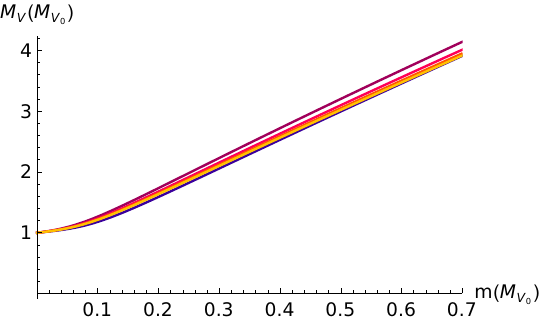}
        \caption{$M_V$}
    \end{subfigure}\hfill
    \begin{subfigure}{0.42\textwidth}
        \centering
        \includegraphics[width=\textwidth]{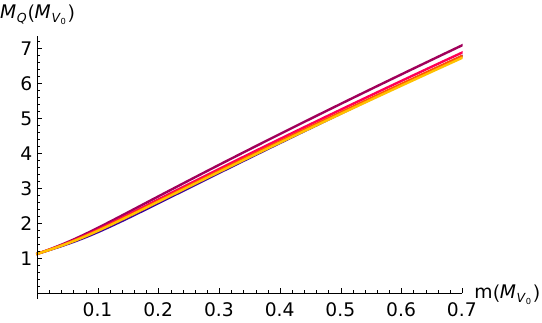}
        \caption{$M_Q$}
    \end{subfigure}\hfill
        \begin{subfigure}{0.42\textwidth}
        \centering
        \includegraphics[width=\textwidth]{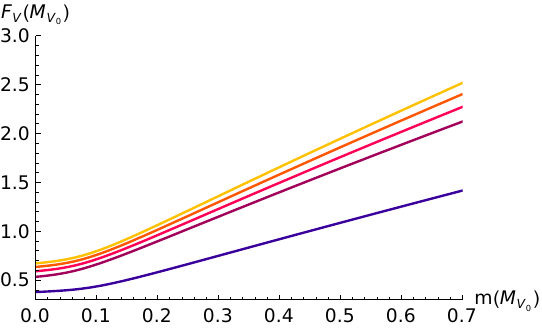}
        \caption{$F_V$}
    \end{subfigure}\hfill
        \begin{subfigure}{0.42\textwidth}
        \centering
        \includegraphics[width=\textwidth]{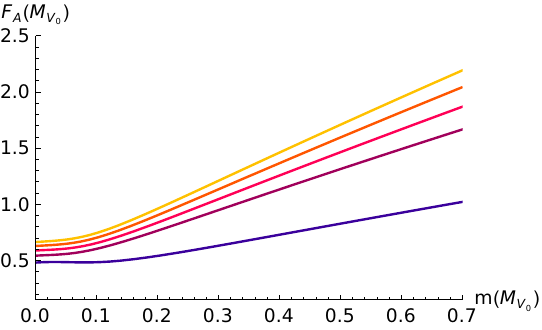}
        \caption{$F_A$}
    \end{subfigure}\hfill
        \begin{subfigure}{0.42\textwidth}
        \centering
        \includegraphics[width=\textwidth]{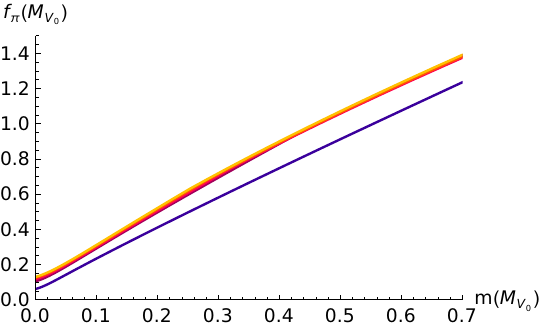}
        \caption{$f_\pi$}
    \end{subfigure}\hfill
    \caption{Bound states' masses and decay constants vary with increasing UV fermion mass, with different gauge groups in the degenerate case when $B=1$. The plots are shown in  units of  $M_{V0}$ at zero fermion mass.}
    \label{fig:deg_flucs_vs_mq}
\end{figure}

\noindent rising. This fall is because the IR value of $L_0$, when a small fermion mass is included, lies a little further from the region of the $\rho-L$ plane where the strongest conformal symmetry breaking is present. The $\sigma$ mass is very sensitive to the rate of the running of $\gamma$.
In figure~\ref{fig:deg_flucs_vs_mq} we show this mass dependence of the spectrum and the decay constants for different gauge group choices.

\section{The theory with mass splitting}\label{sec:non_degenerate_case}
In this section we consider the case where there are two distinct fermion masses in the theory $m_1$ and $m_2$ as in the left hand matrix in eq.~\eqref{masses}. This case shows the power of the non-abelian flavour structure and also that the bulk Higgs mechanism continues to act in this more complicated case. This section also prepares us for our final case where we will enact a similar splitting by a mix of mass terms and NJL interactions.

\subsection{Vacuum}

In particular let's consider $X$ field vevs of the form
\small
\begin{equation}
    \label{eq:nondeg_vac}
    L=\left(
\begin{array}{cccc}
 0 & L_1(\rho) & 0 & 0 \\
 -L_1(\rho) & 0 & 0 & 0 \\
 0 & 0 & 0 & L_6(\rho) \\
 0 & 0 & -L_6(\rho) & 0 \\
\end{array}
\right).
\end{equation}
\normalsize
$L_1$ and $L_6$ will differ by the choice of UV boundary asymptotic so there are two distinct fermion masses. In the case $L_1=L_6$ there are 10 unbroken generators forming an Sp(4) subgroup: the broken generators are, as we have seen, $T^8,T^{10},T^{11},T^{14}, T^{15}, T^{16}$.
When $L_1 \neq L_6$ the unbroken generators are just the six $T^{1-6}$ leaving an unbroken SU(2)$\times$SU(2) sub-group. Note that as we arranged electroweak representations in eq.~\eqref{ewposition}, these still contain the unbroken electroweak gauge group so these theories can still describe composite Higgs models. 

Here in the holographic model, the coordinate is $r^2=\rho^2\mathbbm{1}_4+L^\dagger L=\text{diag}(r_1^2,r_1^2,r_6^2,r_6^2)$, where $r_1^2=\rho^2+L_1^2$, $r_6^2=\rho^2+L_6^2$. The equations for the vaccum vevs are 
\begin{equation}
    \label{eq:sep_vac_eoms}
    \begin{aligned}
        \pdbr{\lun}-\rho \lun\dmq-{4B\over \rho}\lun\lrb{\lun^2-\ldn^2}&=0,\\
        \pdbr{\ldn}-\rho\ldn\dmq-{4B\over \rho}\ldn\lrb{\ldn^2-\lun^2}&=0,\\
    \end{aligned}
\end{equation}
where $\dmq\equiv\dmq\lrmb{\log\sqrt{\rho^2+\half\lrb{\lun^2+\ldn^2}}}$. With slight amendment due to the presence of the $B$ term factors the solutions look like two different choices of the vacuum configurations in figure~\ref{fig:deg_L0_vs_mq}. 

\subsection{Fluctuations --- spectra}

Here, to reflect the symmetry breaking, we parameterise the scalar fluctuations  as
\small
\begin{equation}
\label{eq:sep_mass_X}
X_f= \left( \begin{array}{cccc}
0 & \sigma_1 +i S_1  & Q_2 - \pi_2 + i \pi_1 - i Q_1 & -Q_4 + \pi_4 +i Q_3 - i \pi_3 \\
 - \sigma_1  - iS_1 & 0 & Q_4 + \pi_4 +i Q_3 +i \pi_3 & Q_2 + \pi_2 + i Q_1 + i \pi_1 \\
 \pi_2 - Q_2 +i Q_1-i\pi_1 & -Q_4 - \pi_4 -i Q_3 - i \pi_3 & 0 & \sigma_2 +i S_2  \\
 Q_4-\pi_4 +i \pi_3 - i Q_3 & -Q_2-\pi_2 -i Q_1 -i \pi_1 & -\sigma_2  -i S_2 & 0 
\end{array} \right)\,.
\end{equation}
\normalsize
The spectrum can be assigned to multiplets of SU(2)$\times$SU(2): $\pi_{1,2,3,4}$ are a (2,2) multiplet, $Q_{1,2,3,4}$ are also a (2,2), whilst $\sigma_1, \sigma_2, S_1$ and $S_2$ are singlets. 

The four $\pi_{1-4}$ were Goldstone's (in the bulk even in the massive case where they described pseudo Goldstones in the gauge theory) in the previous mass degenerate case and remain so here. Their equations are
\begin{equation}
    \label{eq:sep_pion_eoms}
    \begin{aligned}
        \pdbr{\phi_j}-{\rho^3\gsq\over 8}\lrb{\lun+\ldn}\invrud\lrb{(\lun+\ldn)\phi_j-2\sqrt{2}\pi_i}&=0,\\ 
        \pdbr{\pi_i}-\rho\dmq\pi_i&\\ \left. \right. ~~~~
        +{\rho M^2\over 2}\invrud\lrb{\pi_i-{1\over2\sqrt{2}}(\lun+\ldn)\phi_j}-{4B\over\rho}(\lun-\ldn)^2\pi_i&=0\\
        i=1,2,3,4, j=14,8,11,10.&
    \end{aligned}
\end{equation}
The bulk Goldstone solutions have $\phi_i=\sqrt{2}$ and $\pi_i=L_1+L_6$ --- the equation becomes the sum of the vacuum equations eq.~\eqref{eq:sep_vac_eoms}. When the UV asymptotics for $L_1$ and $L_6$ include the masses $m_1$ and $m_2$ they do not give $\pi_i \rightarrow 0$ so these solutions are not massless in the gauge theory (those solutions and hence the masses must be found numerically). 
  
 In this case the symmetry breaking in the bulk is greater and we find six additional bulk Goldstone modes eaten by the bulk gauge fields in the Sp(4)$\rar$SU(2)$\times$SU(2) symmetry breaking pattern.  These include the $Q_1-Q_4$ states. Their equations of motion are
\begin{equation}
    \label{eq:sep_Q_eoms}
    \begin{aligned}
        \pdbr{\phi_j}+{\rho^3 \gsq \over 4}\lrb{\lun-\ldn}\invrud\lrb{\sqrt{2}Q_i-{\lun-\ldn \over 2}\phi_j}&=0,\\
        \pdbr{Q_i}+{\rho^3 M^2\over 2}\invrud Q_i-\rho\dmq Q_i-{4 B\over \rho}\lrb{\lun+\ldn}^2Q_i&\\
        -{\rho^3\over 4\sqrt{2}}\lrb{\lun-\ldn}M^2\invrud\phi_j&=0,\\
        i=1,2,3,4 \text{ for }j=13,7,12,9&.
    \end{aligned}
\end{equation}
The Goldstone modes in the bulk are $\phi_i=\sqrt{2}$, $Q_i=L_1-L_6$ and we obtain the difference of the vacuum equations eq.~\eqref{eq:sep_vac_eoms}.

In addition, the former $S$ and $\pi_5$ states combine into the $S_2=S+\pi_5$ and $S_1=S-\pi_5$ states, which are eaten by $\half(A_{15}\pm A_{16})$ which correspond to  the generators ${1\over 2\sqrt{2}}\text{diag}(1,1,0,0)$ and ${1\over 2\sqrt{2}}\text{diag}(0,0,1,1)$ respectively.
Their equations are 
\begin{equation}
    \label{eq:sep_S1_16_eoms}
    \begin{aligned}
        \pdbr{S_1}-\rho\dmq S_1+{\rho^3M^2\over r_1^4}\lrb{S_1-{\lun\over\sqrt{2}}\phi_{16-15}}-{4B\over\rho}\lrb{\lun^2-\ldn^2}S_1&=0,\\
        \pdbr{\phi_{16-15}}+{\rho^3 L_1 g_5^2\over r_1^4}\lrb{\sqrt{2}S_1-L_1\phi_{16-15}}&=0,
    \end{aligned}
\end{equation}
\begin{equation}
    \label{eq:sep_S2_15_eoms}
    \begin{aligned}
        \pdbr{S_2}-\rho\dmq S_2+{\rho^3M^2\over r_6^4}\lrb{S_2-{\ldn\over\sqrt{2}}\phi_{15+16}}-{4B\over\rho}\lrb{\ldn^2-\lun^2}S_2&=0,\\
        \pdbr{\phi_{15+16}}+{\rho^3 L_6 g_5^2\over r_6^4}\lrb{\sqrt{2}S_2-L_6\phi_{15+16}}&=0.
    \end{aligned}
\end{equation}
The bulk Goldstone modes are $\phi_i = \sqrt{2}$ with $S_1=L_1$ and $S_2=L_6$.

Finally there are two uneaten components of $X$ --- the  scalar fluctuations around the vacua $\lun$ and $\ldn$, which we have called $\s_1$ and $\s_2$, and have mixed equations of motion 
\begin{equation}
    \label{eq:sep_Sud_eoms}
    \begin{aligned}
        \pdbr{\s_1}+\lrb{{\rho^3 M^2\over r_1^4 }-{ 4B\over \rho}\lrb{3\lun^2-\ldn^2}-\rho\dmq-{\rho\lun^2\over2 r_{arg,16}^2}\dmqp}\s_1&\\
         +\lun\ldn\lrb{{8B\over \rho}-{\rho\over 2r_{arg,16}^2}\dmqp}\s_2&=0,\\
        \pdbr{\s_2}+\lrb{{\rho^3 M^2\over r_6^4 }-{4 B\over \rho}\lrb{3\ldn^2-\lun^2}-\rho\dmq-{\rho\ldn^2\over 2r_{arg,16}^2}\dmqp}\s_2&\\
        +\lun\ldn\lrb{{8B\over \rho}-{\rho\over2 r_{arg,16}^2}\dmqp}\s_1&=0,\\ 
    \end{aligned}
\end{equation}
where $\dmq\equiv\dmq\lrb{\log r_{arg,16}}$, $\dmqp={d \dmq\over d\log(r_{arg,16})}$,
$r_{arg,16}^2=\rho^2+\half\lrb{\lun^2+\ldn^2}$.
In deriving the equations of motion, we used
\begin{equation}
    \label{eq:sep_sub_rule}
    \begin{aligned}
        A+2\lrb{\lun^2+\ldn^2} {d A\over d \Tr(\rho^2\mathbbm{1}_4+L^\dagger L)}&=\dmq\lrb{\log r_{arg,16}},\\
       {d A\over d \Tr(\rho^2\mathbbm{1}_4+L^\dagger L)}+\lrb{\lun^2+\ldn^2}{d^2 A\over d \Tr(\rho^2\mathbbm{1}_4+L^\dagger L)^2}&={1\over16 r_{arg}^2} \dmqp.
    \end{aligned}    
\end{equation}
The  transverse pieces of the sixteen gauge bosons then generate equations of motion associated with the six unbroken vectors --- these states are a (3,1) and a (1,3) and the ten broken axial vectors --- two (2,2) and two singlets
\begin{equation}
    \label{eq:sep_eoms}
    \begin{aligned}
        \pdbr{A_i}+{\rho^3 M_{A_i}^2\over r_j^4}A_i&=0,\:\\
        i=(1\pm 4,2\pm 5,3\pm 6),j&=1,6, \\
        \pdbr{A_i}+{\rho^3\over8}\invrud\lrmb{4M_{A_i}^2-(\lun-\ldn)^2\gsq}A_i&=0,\:i=7,9,12,13,\\
        \pdbr{A_i}+{\rho^3\over8}\invrud\lrmb{4M_{A_i}^2-(\lun+\ldn)^2\gsq}A_i&=0,\:i=8,10,11,14,\\
        \pdbr{A_i}+{\rho^3\over r_j^4}\lrmb{M_{A_i}^2-L_j^2\gsq}A_i&=0,\:i=16\pm15,j=6,1.
    \end{aligned}
\end{equation}
\begin{figure}
    \centering
    \begin{subfigure}{0.5\textwidth}
        \centering
        \includegraphics[width=\textwidth]{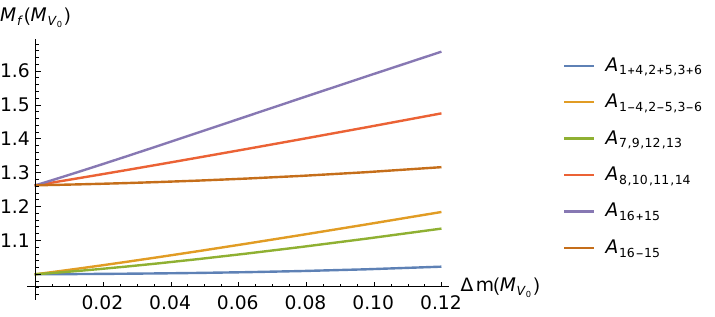}
        \caption{$M_f$, $f$ = spin 1 state, Sp(4)}
    \end{subfigure}\hfill
    \begin{subfigure}{0.45\textwidth}
        \centering
        \includegraphics[width=\textwidth]{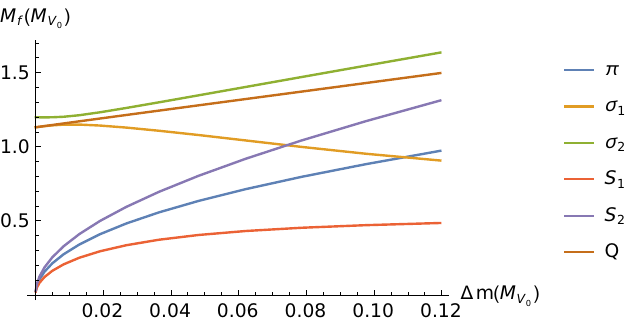}
        \caption{$M_f$, $f$ = spin 0 state, Sp(4)}
    \end{subfigure}\hfill
\caption{\label{fig:nondeg_flucs_vs_mq_Sp4} 
    Dependence of meson masses on the difference $\Delta m = m_2 - m_1$ of the UV fermion masses for Sp(4) gauge groups in units of $M_{V_0}$. We have set $B=1$ and $m_1=0.014 M_{V_0}$. The $A_{1+4,2+5,3+6}$ and $A_{1-4,2-5,3-6}$ vector states are members of the (3,1) and (1,3); $A_{7,9,12,13}$ and $A_{8,10,11,14}$ are members of the two (2,2); $A_{16+15}$ and $A_{16-15}$ are singlets.}
\end{figure}

\begin{figure}
    \centering
    \begin{subfigure}[b]{0.5\textwidth}
        \centering
        \includegraphics[width=\linewidth]{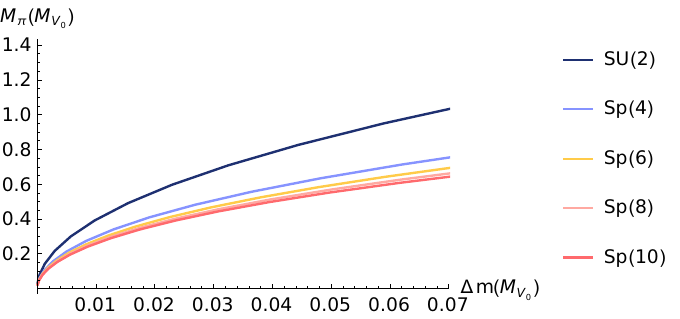}
        \caption{$M_\pi$}
    \end{subfigure}\hfill
    \begin{subfigure}{0.42\textwidth}
        \centering
        \includegraphics[width=\textwidth]{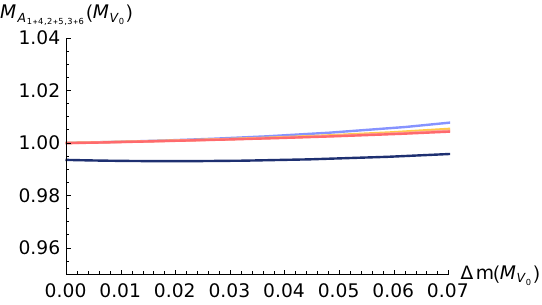}
        \caption{$M_{A_{1+4,2+5,3+6}}$}
    \end{subfigure}\hfill
    \begin{subfigure}{0.3\textwidth}
        \centering
        \includegraphics[width=\textwidth]{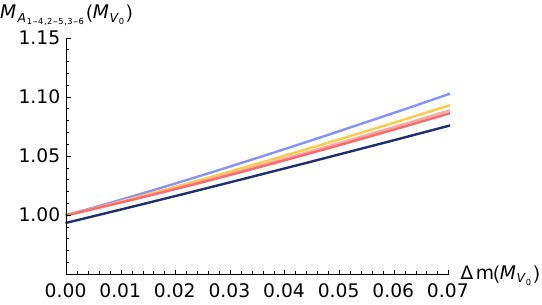}
        \caption{$M_{A_{1-4,2-5,3-6}}$}
    \end{subfigure}\hfill
    \begin{subfigure}{0.3\textwidth}
        \centering
        \includegraphics[width=\textwidth]{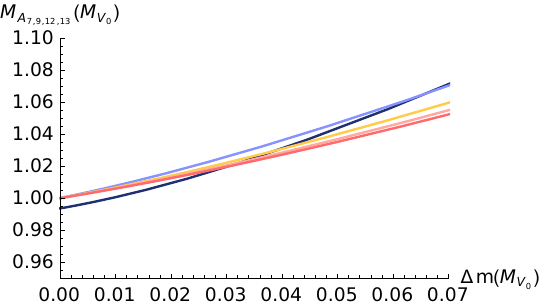}
        \caption{$M_{A_{7,9,12,13}}$}
    \end{subfigure}\hfill
    \begin{subfigure}{0.3\textwidth}
        \centering
        \includegraphics[width=\textwidth]{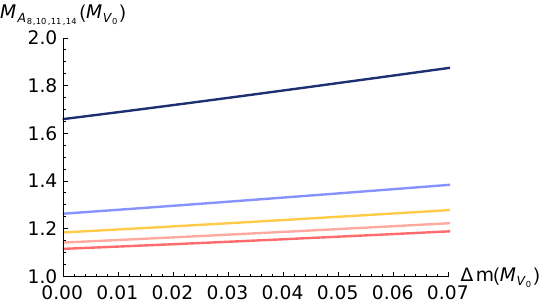}
        \caption{$M_{A_{8,10,11,14}}$}
    \end{subfigure}\hfill
    \begin{subfigure}{0.3\textwidth}
        \centering
        \includegraphics[width=\textwidth]{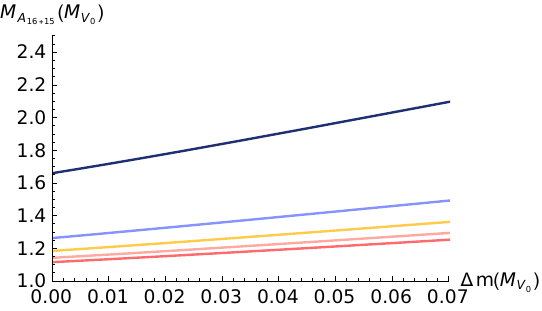}
        \caption{$M_{A_{16+15}}$}
    \end{subfigure}\hfill
    \begin{subfigure}{0.3\textwidth}
        \centering
        \includegraphics[width=\textwidth]{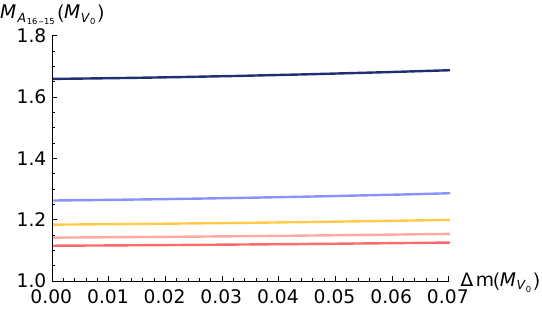}
        \caption{$M_{A_{16-15}}$}
    \end{subfigure}\hfill
    \begin{subfigure}{0.3\textwidth}
        \centering
        \includegraphics[width=\textwidth]{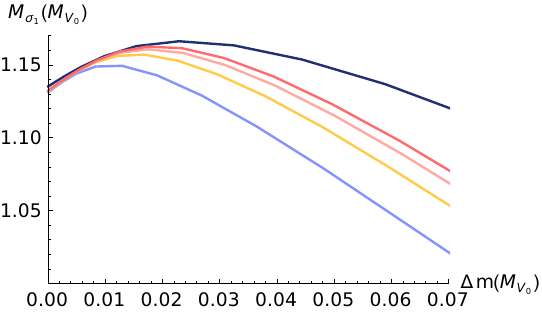}
        \caption{$M_{\s_1}$}
    \end{subfigure}\hfill
    \begin{subfigure}{0.3\textwidth}
        \centering
        \includegraphics[width=\textwidth]{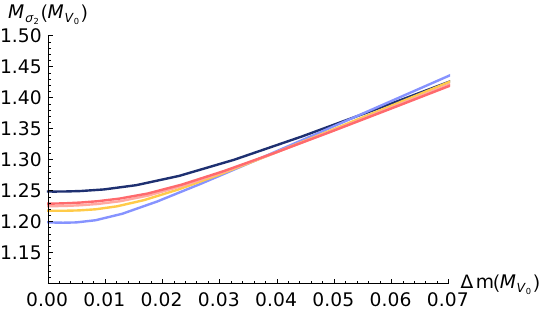}
        \caption{$M_{\s_2}$}
    \end{subfigure}\hfill
    \begin{subfigure}{0.3\textwidth}
        \centering
        \includegraphics[width=\textwidth]{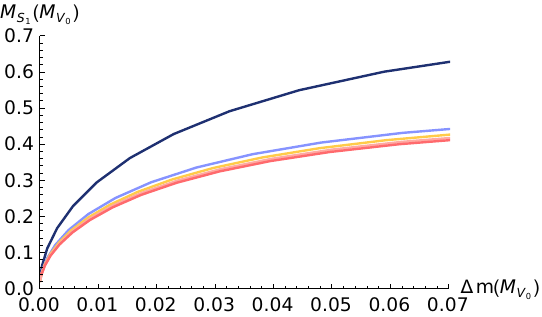}
        \caption{$M_{S_1}$}
    \end{subfigure}\hfill
    \begin{subfigure}{0.3\textwidth}
        \centering
        \includegraphics[width=\textwidth]{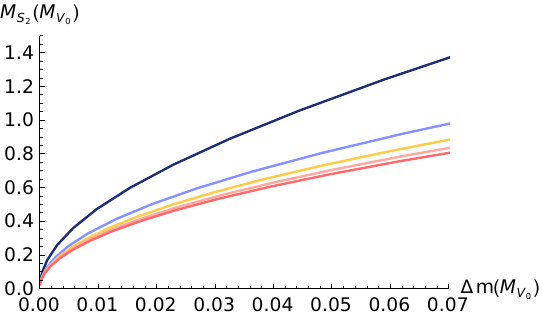}
        \caption{$M_{S_2}$}
    \end{subfigure}\hfill
    \begin{subfigure}{0.3\textwidth}
        \centering
        \includegraphics[width=\textwidth]{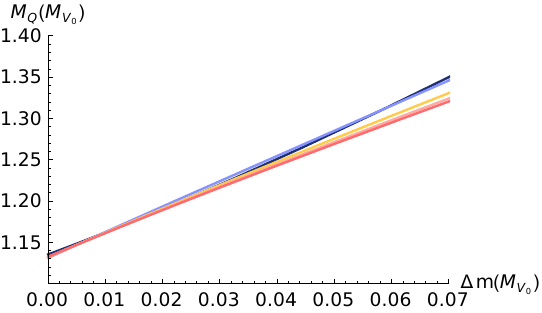}
        \caption{$M_{Q}$}
    \end{subfigure}\hfill
    \caption{Dependence of meson masses on the difference $\Delta m = m_2 - m_1$ of the UV fermion masses for different gauge groups in unit $M_{V_0}$. We have set $B=1$ and $m_1=0.014 M_{V_0}$.}
    \label{fig:nondeg_flucs_vs_mq}
\end{figure}

\begin{figure}
    \centering
    \begin{subfigure}[b]{0.5\textwidth}
        \centering
        \includegraphics[width=\linewidth]{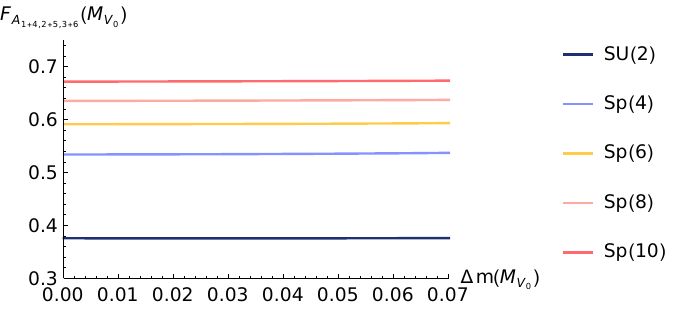}
        \caption{$F_{A_{1+4,2+5,3+6}}$}
    \end{subfigure}\hfill
    \begin{subfigure}{0.42\textwidth}
        \centering
        \includegraphics[width=\textwidth]{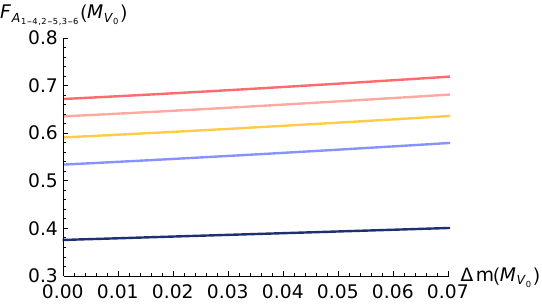}
        \caption{$F_{A_{1-4,2-5,3-6}}$}
    \end{subfigure}\hfill
    \begin{subfigure}{0.3\textwidth}
        \centering
        \includegraphics[width=\textwidth]{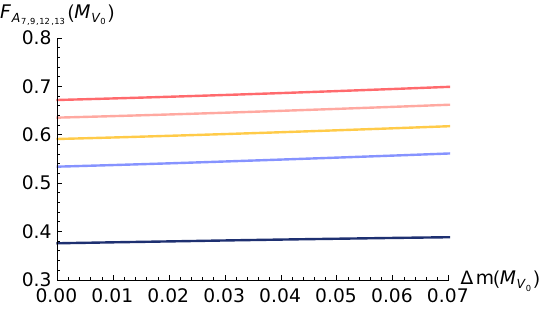}
        \caption{$F_{A_{7,9,12,13}}$}
    \end{subfigure}\hfill
    \begin{subfigure}{0.3\textwidth}
        \centering
        \includegraphics[width=\textwidth]{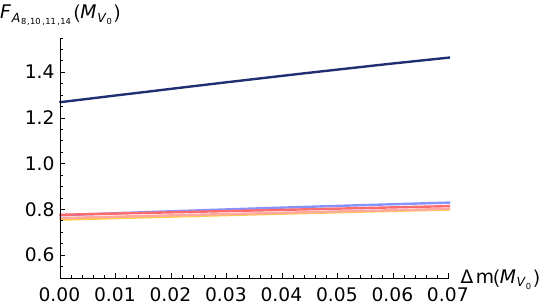}
        \caption{$M_{A_{8,10,11,14}}$}
    \end{subfigure}\hfill
    \begin{subfigure}{0.3\textwidth}
        \centering
        \includegraphics[width=\textwidth]{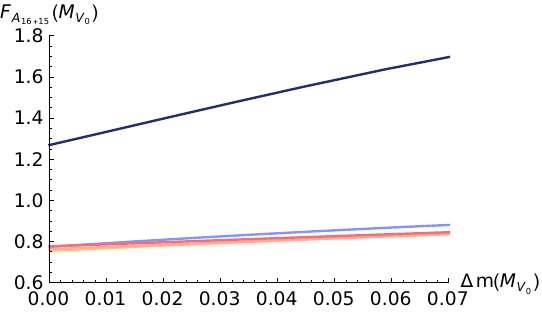}
        \caption{$M_{A_{16+15}}$}
    \end{subfigure}\hfill
    \begin{subfigure}{0.3\textwidth}
        \centering
        \includegraphics[width=\textwidth]{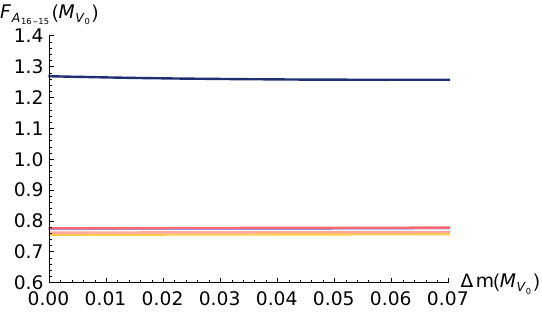}
        \caption{$M_{A_{16-15}}$}
    \end{subfigure}\hfill
    \begin{subfigure}{0.3\textwidth}
        \centering
        \includegraphics[width=\textwidth]{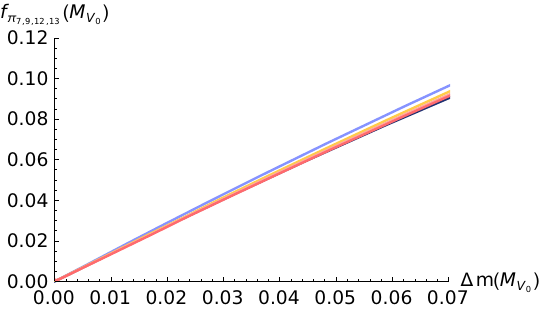}
        \caption{$f_{\pi_{7,9,12,13}}$}
    \end{subfigure}\hfill
    \begin{subfigure}{0.3\textwidth}
        \centering
        \includegraphics[width=\textwidth]{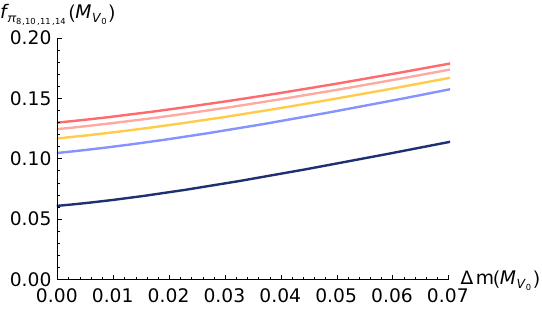}
        \caption{$f_{\pi_{8,10,11,14}}$}
    \end{subfigure}\hfill
    \caption{Meson decay constants vary with increasing UV fermion mass difference, with different gauge groups in the non-degenerate case. We have set $B=1$ and $m_1=0.014$.}
    \label{fig:nondeg_f_vs_mq}
\end{figure}

 As a first example, we have computed the mass spectrum and the decay constants for $B=1$ when $m_1=0.014$ and for varying $m_2$ for the Sp(4) theory. 
We plot the masses with increasing UV fermion mass differences in figure~\ref{fig:nondeg_flucs_vs_mq_Sp4}. As one would expect, the generic trend is for the fields made of the fermion whose mass we are varying to rise as the fermion mass does. The separation of the masses of the different multiplets is apparent as the fermion mass degeneracy is lost. We note that the axial vectors $A_{7,9,12,13}$ become vector states in the limit $L_1 \to L_6$. The pseudo-Goldstone modes show Gell-Mann-Oakes-Renner behaviour with the heavier fermion mass since the lowest mass is small here. The $\sigma_1$ mass again shows some slight drop as it responds to the presence of the higher second UV mass in the model but at very large UV masses its mass must become independent of the larger UV mass. In figure~\ref{fig:nondeg_flucs_vs_mq} and figure~\ref{fig:nondeg_f_vs_mq} we show for completeness more detailed results of the $N_c$ dependence of the masses and decay constants of the theories with split masses.

\subsection{First phenomenological considerations}

There are two scenarios where our results
can be applied: (i) one interprets the
pNGBs as the multiplet containing the Higgs 
bosons, see e.g.~\cite{Ferretti:2013kya,Cacciapaglia:2014uja,Ferretti:2016upr,Belyaev:2016ftv,BuarqueFranzosi:2016ooy} or (ii) one interprets it
as a dark sector model to explain the observed dark matter relic density
\cite{Hochberg:2014kqa,Kulkarni:2022bvh,Zierler:2022uez}.

We start with option (i). Our calculation
of the pseudo-Nambu-Goldstone bosons (pNGB) mass does not take into account
the correction due to the electroweak
gauge gauge couplings and the top Yukawa coupling.  Thus, we cannot set its mass to 125~GeV but we still need these states to have masses of order 100 GeV to construct a sensible electroweak model. In ref.~\cite{BuarqueFranzosi:2016ooy} 
bounds on the masses of the spin-1 resonances 
have been obtained of up to 2 TeV assuming 
that (i) they are practically mass degenerate
and (ii) certain values of the couplings.
This is at first glance a rather strong bound and is a consequence of the mixing between the
usual vector bosons and some of the spin-1
resonances. We give in table~\ref{tab:dict_vectors} a dictionary between our notation and the one of
ref.~\cite{BuarqueFranzosi:2016ooy}.
\begin{table}[t]
 \centering
 \begin{tabular}{cccc} \hline
  $A_{1+4,2+5,3+6}$ & $A_{1-4,2-5,3-6}$ & $A_{7,9,12,13}$ & $A_{8,10,11,14}$\\
  $v^{0,\pm}$ & $s^{0,\pm}$ & $\tilde s^{0,\pm}$ + $\tilde v^0$ & $a^{0,\pm}$ + $x^0$ \\ \hline 
    \end{tabular}
 \caption{Dictionary between our notation,
 upper row, and the one of ref.~\cite{BuarqueFranzosi:2016ooy} for
 the spin-1 states. 
Our states $A_{16+15}$  and $A_{16-15}$ correspond to an admixture of $\tilde x^0$ of ref.~\cite{BuarqueFranzosi:2016ooy} and the additional U(1) axial vector boson.}
\label{tab:dict_vectors}
\end{table}
While these bounds depend on certain details,
it is clear that the combination of our results
in figure~\ref{fig:nondeg_flucs_vs_mq} with experimental data
require that $M_{V_0}/M_\pi \gsim 10$ (note the analysis of ref.~\cite{BuarqueFranzosi:2016ooy} is based on a luminosity of about 3 fb$^{-1}$ only, thus the real bounds will actually be stronger. However, this does not affect our conclusions).
This in turn immediately implies that the
UV fermion masses cannot be much larger than
$\sim 10^{-3} M_{V_0}$. Therefore, this model
predicts that there should be two additional
gauge singlet states with masses close to
$m_H$ which is still compatible with present
data. Moreover, 
the lowest lying bosonic states are 
$A_{1+4,2+5,3+6},A_{1-4,2-5,3-6}$ and $A_{7,9,12,13}$ followed by $Q$, $S_1$, $S_2$ and $A_{8,10,11,14}, A_{16+15}, A_{16-15}$ with a mass
ratio of about 
\begin{align}
M_{A_{1+4},A_{1-4},A_{7}} : M_{Q,S_1} : M_{S_2} :
M_{A_{8},A_{16+15},A_{16-15}} \simeq 1 : 1.1 : 1.2 : 1.26\,.
\end{align}
where we have combined the states which are
nearly mass degenerate. This immediately
implies that the scalars $Q$ and $S_{1,2}$
will not be observed at the high luminosity 
LHC as their production cross sections are rather low for masses above 2 TeV.

We now turn to option (ii), namely dark matter from decoupled strongly interacting sectors \cite{Hochberg:2014kqa}. Scenarios in which only the pNGBs contribute to the relic density are
highly constraint and potentially already excluded \cite{Bernreuther:2023kcg}. 
However, it has also been shown in  ref.~\cite{Bernreuther:2023kcg} that the constraints can be relaxed if additional states are present, in particular they
considered additional vector states $V$ with $1.4\, M_\pi \lsim M_V < 2 \,M_\pi$ in 
a model based on an SU(N) gauge group. We see from
figure~\ref{fig:deg_all_flucs_vs_mq} that such a scenario can be realized for
$0.02 M_{V_0} \lsim m \lsim 0.1 M_{V_0}$. Note, that in our model  $\sigma$ 
is also present in such a scenario which will
additionally contribute to the calculation of
the relic dark matter density which
will be investigated in a future work.
In the case of a mass splitting $m_1\ne m_2$
the details of such a scenario will change but the main overall features
are the same as in the degenerate
case as can be seen in 
figure~\ref{fig:nondeg_flucs_vs_mq_Sp4}.
Note that having $2 M_{S_1} < M_{A_{1+4}}$
does not invalidate the assumptions
of \cite{Bernreuther:2023kcg}
because the $A_{1+4}$ does not decay into
$2 S_1$. The reason is, that $S_1$ is
a singlet of the remaining SU(2)$_D$
flavour symmetry whereas $A_{1+4,2+5,3+6}$ is
a triplet.

\section{Competition between a mass and the NJL interaction}
\label{sec:njl}

We now turn to the main goal of our project. We will study the SU(2) gauge theory with both a mass term and NJL interactions that each favours a different vacuum and look at the transition between the two physics regimes. 

In particular we will start from the theory described in section \ref{sec:degenerate_case} with a single, small, common mass between each of two of the four spinors. The vacuum eq.~\eqref{1stvacnow} aligns with the mass term and preserves an Sp(4) global symmetry. There are 6 light pseudo-Goldstones $\pi_i$ with $i=1..5$, $S$ in the basis eq.~\eqref{fluc1} including a four-plet that is a bi-fundamental of SU(2)$_L \times$ SU(2)$_R$ that can be identified with the standard model Higgs in composite Higgs models.

To this case we will add the four-fermion operators eq.~\eqref{NJLops} that link left and right handed fields. These operators favour the condensation of $Q_4$ whose vev will switch on at a second order transition above some critical NJL coupling. The $Q_4$ vev will  place the theory in a vacuum of the form in eq.~\eqref{Xch}.

\subsection{Holographic vacuum}

In the holographic model our $X$ field will take the form
\small
\begin{equation}
    \label{eq:njl_vac}
    X=\left(
    \begin{array}{cccc}
     0 & L_0(\rho) & 0 &  -Q(\rho) \\
     -L_0(\rho) & 0 & Q(\rho) & 0 \\
     0 & -Q(\rho) & 0 & L_0(\rho) \\
     Q(\rho) & 0 & -L_0(\rho) & 0 \\
    \end{array}
    \right).
\end{equation}
\normalsize

It is useful to note that one can use the unitary transformation in eq.~\eqref{trans1} to recast the
vacuum to
\small
\begin{equation}  \label{firstrotate}
U^T X U =
\left(
    \begin{array}{cccc}
     0 & L_0+Q & 0 &  0 \\
     -L_0- Q & 0 & 0 & 0 \\
     0 & 0 & 0 & L_0 -Q\\
     0 & 0 & -L_0+Q & 0 \\
    \end{array} 
    \right)  =
    \left(
    \begin{array}{cccc}
     0 & L_p & 0 &  0 \\
     -L_p & 0 & 0 & 0 \\
     0 & 0 & 0 & L_m\\
     0 & 0 & -L_m & 0 \\
    \end{array} 
    \right),
\end{equation}
\normalsize
with $L_{p/m} = L_0 \pm Q$. The vacuum now takes the form we had for the mass split case in eq.~\eqref{eq:nondeg_vac}. 

There is one subtlety we will encounter. As the NJL coupling rises one reaches a point where the UV mass in $Q$ induced by the NJL interaction becomes equal to the bare fermion mass in $L_0$. At this point the solutions have $Q=L_0$ at all $\rho$ (the NJL solutions are just the $L_0$ solutions reinterpreted so this is inevitable). For larger Q naively $L_m$ becomes negative and in terms of figure~\ref{fig:deg_L0_vs_mq} lies below the axis. Here though one must remember that one can make a phase rotation on the solution and return it to living above the axis. This will be the preferred vacuum which one can see if one returns to the original string construction and thinks of the mixed ``m-p'' bound states as strings between the two embeddings. The length of the string and hence the mass of the bound states is minimized if they are nearer which happens when both solutions are above the axis. The conclusion is that when $Q>L_0$ one should use the transformation from eq.~\eqref{trans2} rather than that in eq.~\eqref{trans1} to place the vacuum in the state
\small
\begin{equation} 
U^T X U =
\left(
    \begin{array}{cccc}
     0 & L_0+Q & 0 &  0 \\
     -L_0- Q & 0 & 0 & 0 \\
     0 & 0 & 0 & -L_0 +Q\\
     0 & 0 & L_0-Q & 0 \\
    \end{array} 
    \right)  =
    \left(
    \begin{array}{cccc}
     0 & L_p & 0 &  0 \\
     -L_p & 0 & 0 & 0 \\
     0 & 0 & 0 & -L_m\\
     0 & 0 & L_m & 0 \\
    \end{array} 
    \right). 
\end{equation}
\normalsize
Another sensible crosscheck is that in the limit $L_0 \ll Q$ we have returned to the form eq.~\eqref{1stvacnow} but now with $L_0=Q$. 

Since the vacuum can be placed in the form of that for the split mass solutions the vacuum naively breaks U(4) to SU(2)$\times$ SU(2). Explicitly,
$L_0(\rho)$ breaks the generators $T^8$ (with associated Goldstone $\pi_2$), $T^{10} (\pi_4)$, $T^{11}(\pi_3)$, $T^{14} (\pi_1)$, $T^{15} (\pi_5)$, $T^{16} (S)$, whilst $Q(\rho)$ breaks $T^4 (Q_1)$, $T^5 (Q_2)$, $T^6 (Q_3)$, $T^{9} (Q_5)$, $T^{11} (S)$, $T^{16}(\pi_3)$. This leaves $T^{1,2,3}$ that are the custodial SU(2)$_V$ of the electroweak model and a second SU(2) with generators $T^{7,12,13}$. In fact this second SU(2) is explicitly broken by the NJL interactions --- we will show this clearly when we discuss the boundary conditions on the fluctuation fields below.

To formulate the equations for $L_p$ and $L_m$ we note
that after the unitary rotation, the radial direction is now $r^2=\text{diag}(r_p^2,r_p^2,r_m^2,r_m^2)$, where $r_p^2=\rho^2+(L_0+Q)^2$, $r_m^2=\rho^2+(L_0-Q)^2$.

The vacua's equations of motion are 
\begin{equation}
    \label{eq:njl_vac_eoms}
    \begin{aligned}
        \pdbr{\lpn}-\rho\dmq\lpn-{4B(\lpn^2-\lnn^2)\over \rho}\lpn&=0,\\
        \pdbr{\lnn}-\rho\dmq\lnn-{4B(\lnn^2-\lpn^2)\over \rho}\lnn&=0,\\
    \end{aligned}
\end{equation}
where $\dmq=\dmq\lrmb{\log\lrb{\sqrt{\rho^2+\half(\lpn^2+\lnn^2)}}}$. 
 $L_0$ and $Q$ represent the vacuum that gives the fermion mass $m$ and the NJL contributions respectively. In this basis, the embeddings' equations are
 \begin{equation}
     \label{eq:njl_vac_LQ_eoms}
         \begin{aligned}
        \pdbr{L_0}-\rho\dmq L_0-{16B Q^2\over \rho}L_0&=0,\\
        \pdbr{Q}-\rho\dmq Q-{16B L_0^2\over \rho}Q&=0.\\
    \end{aligned}
 \end{equation}
 In the numerical computations, we keep the notation $\lpn$ and $\lnn$. More precisely, their UV behaviours are
\begin{equation}
    \label{eq:njl_vac_UV}
    L_p(\rho_{UV})\sim \mcl{J}+m+ {\mcl{O}\over \rho^2}+{c_0\over\rho^2},\quad L_m(\rho_{UV})\sim m-\mcl{J}+{c_0\over\rho^2}- {\mcl{O}\over \rho^2}.
\end{equation}
$m$ is UV fermion mass, $c_0$ is the corresponding vev. $\mcl{J}$ is the source term at the boundary that is introduced by the higher dimensional operator 
\begin{equation}
    \label{eq:njl_source}
    \mcl{L}_{UV}=G\mcl{O}^\dagger\mcl{O},\quad \mcl{J}=G\langle \mcl{O}^\dagger \rangle,
\end{equation}
where 
\begin{equation}
    \label{eq:njl_scalar_G}
    \quad G={g_s^2\over\L_{UV}^2}\quad \text{or}\quad g_s^2={\mcl{J}\L_{UV}^2\over \langle \mcl{O}^\dagger \rangle}.
\end{equation}
$\L_{UV}$ is the UV cut-off. Note that $\mcl{J}=m_{UV}=0$ when $g_s^2=0$ or below the critical value. For $g_s^2$ greater than the critical value, one by hand includes a UV value $m_{UV}$ and reads off the vev and the source by 
\begin{equation}
    \label{eq:njl_readout}
    {Q'(\L_{UV})\over-2\L_{UV}^{-3}}=\langle \mcl{O}^\dagger \rangle,\quad \mcl{J}=m_{UV}-{\langle \mcl{O}^\dagger \rangle\over \L_{UV}^2}.
\end{equation}

\begin{figure}
    \centering
    \includegraphics[width=\linewidth]{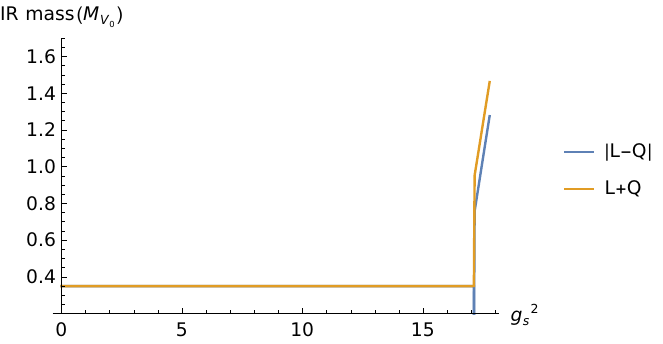}
    \caption{The vacuum of the SU(2) gauge theory with NJL interactions. IR masses  in units of the vector mass $M_{V_0}$ are plotted against the scalar NJL coupling $g_s^2$, $B=0.1$, $m=0.12\, M_{V_0}$, $\L_{UV}=12.1 M_{V_0}$. Initially we plotted the IR masses of $L+Q$ and $L-Q$. As soon as $Q>L$, we flip the $L-Q$ and show the IR mass of $Q-L$.  }
    \label{fig:njl_vac}
\end{figure}

This gives a sufficiently good approximation to the UV behaviour, as one can see this by plotting the UV asymptote on top of the fluctuation's profile. In the $L_{p,m}$ basis, we imposed the following boundary conditions to solve the embbedings
\begin{equation}
    \label{eq:NJL_vac_bcs}
    \begin{aligned}
      & L_p(\rho_{IR,p})=\rho_{IR,p},\quad L_p'(\rho_{IR,p})=0,\quad L_p(\L_{UV})=m_{UV,p},\\
       &\begin{cases}
           L_m(\rho_{IR,m})=-\rho_{IR,m},\, L_m(\L_{UV})=m_{UV,m},\,L_m(\rho)\equiv -L_m(\rho),&m_{UV,m}<0\\
           L_m(\rho_{IR,m})=\rho_{IR,m},\, L_m(\L_{UV})=m_{UV,m},&m_{UV,m}>0
       \end{cases},\\ &L_m'(\rho_{IR,m})=0.\\
    \end{aligned}
\end{equation}
The boundary conditions for $L_m=L_0-Q$ requires the NJL source term is smaller than the fermion mass $m$. At the point $L_0(UV)=Q(UV)$, $L_m=-L_m=0$, one must, as described above, jump from the vacuum where $L_m=L_0-Q=0$ to $-L_m=Q-L_0=0$.  We thus flip the solution above the $\rho$-axis to get $Q-L_0$. We show the IR masses of the embeddings $L_0\pm Q$ for an SU(2) gauge group in figure~\ref{fig:njl_vac} where $L_0(UV)$ is fixed to 0.12 in unit $M_V(m=0)=1$, and we take $B=0.1$. The results show the expected second order transition at a critical $g_s^2$ value near $g_s^2=17$. The NJL interaction naturally wants to raise the $Q$ vev to the cut off scale we chose $\Lambda_{UV}= 12.1 M_{V_0}$ so the rise is sharp. This means that the rotation from a composite Higgs model behaviour to a pure technicolour model behaviour is very fine tuned in $g_s^2$.

\subsection{Fluctuations}

In the previous sub-section we demonstrated how to find the vacuum functions $L_0$ and $Q$ in eq.~\eqref{eq:njl_vac} as a function of the NJL coupling $g_s$. We now turn to the fluctuation masses in those vacua. Naively this is straightforward: as we have seen using eq.~\eqref{trans1} or eq.~\eqref{trans2} we can always place the vacuum vev in the form of the split mass case of section~\ref{sec:non_degenerate_case} . The fluctuation equations are therefore those of section~\ref{sec:non_degenerate_case}. 

However, we must be careful to track the scalar operators that see the NJL interaction through the UV boundary conditions on their bulk dual scalar. In particular in the basis eq.~\eqref{eq:njl_vac} if we write the fluctuations as in eq.~\eqref{fluc1} then we can clearly see that the fluctuations $Q_{1,2,3,4}$ correspond to operators that directly see the NJL interaction.  Note $\pi_{1,2,3,4}$ do not see the NJL operator in their boundary conditions because they are the phases of the $Q_i$.

In particular these four fields must have the UV boundary conditions 
\begin{equation}
    Q_i \simeq {\cal J} + {{\cal O} \over \rho^2}\quad
{\rm with}\quad
    {\cal J} = {g_s^2 \over \Lambda^2} {\cal O}, 
\end{equation}
where $g_s^2$ is the value of the NJL coupling already found for the vacuum solution under consideration. Note that below the critical value of the coupling $g_s^2$, the vacuum solutions have $Q=0$ but one still needs to apply the NJL condition to the fluctuations at a particular value of $g_s^2$. 

Now when we make the basis rotation in eq.~\eqref{trans1} the fields ``move" in the matrix to the positions
\footnotesize
\begin{equation} \label{flucnew}
X_f= \left( \begin{array}{cccc}
0 & \sigma +Q_4 +i S +i \pi_3 & -Q_2 - \pi_2 - i \pi_1 - i Q_1 & Q_5 + \pi_4 +i Q_3 +i \pi_5 \\
 - \sigma-Q_4 -iS -i \pi_3  & 0 & -Q_5 + \pi_4 +i Q_3 -i \pi_5 & -Q_2 + \pi_2 + i Q_1 - i \pi_1 \\
 \pi_2 + Q_2 +i Q_1+i\pi_1 & Q_5 - \pi_4 -i Q_3 + i \pi_5 & 0 & \sigma - Q_4 +i S - i \pi_3 \\
 -Q_5-\pi_4 -i \pi_5 - i Q_3 & Q_2-\pi_2 -i Q_1 +i \pi_1 & -\sigma + Q_4 -i S + i \pi_3 & 0 
\end{array} \right),
\end{equation}
\normalsize 
in the basis where the vev is that in eq.~\eqref{firstrotate}. Thus there is a relabelling of the fields that must be accounted for if we are to use the equations from section~\ref{sec:non_degenerate_case}.

\begin{figure}
    \centering   
    \begin{subfigure}{0.5\textwidth}
    \includegraphics[width=\textwidth]{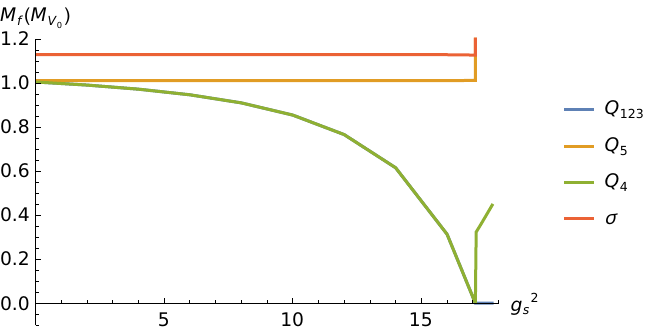}
        \caption{$M_f^2$, $f=Q_{123}, Q_4, Q_5,\sigma$} 
        \label{fig:njl_flucs_vs_gssq_vec}
    \end{subfigure}\hfill
    \begin{subfigure}{0.5\textwidth}
        \centering
    \includegraphics[width=\textwidth]{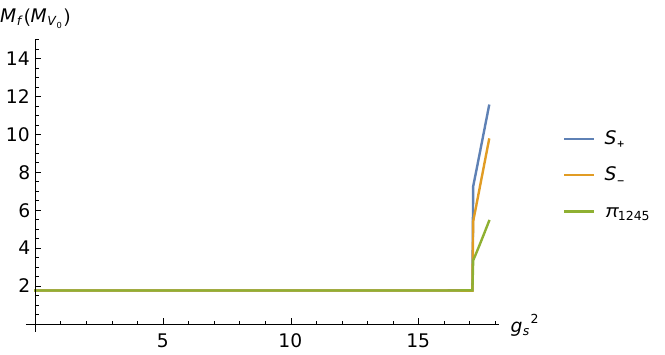}
        \caption{$M_f^2$, $f=\pi_{1245}, S_+, S_-$}
        \label{fig:njl_flucs_vs_gssq_scalar}
    \end{subfigure}\hfill
    \caption{Scalar bound states' masses vary with increasing scalar NJL coupling $g_s^2$, with gauge group SU(2), $B=0.1$, $m_L=0.12M_{V_0}$, $\L_{UV}=12.1 M_{V_0}$.} 
    \label{fig:njl_flucs_vs_gssq}
\end{figure}

There is an important piece of physics in this relabelling. Whilst in the split mass case equations eq.~\eqref{eq:sep_Q_eoms} the four $Q_i$ are in that case a (2,2) degenerate multiplet of SU(2)$\times$ SU(2) here those $Q_i$ contain $Q_{1,2,3}$ and $Q_5$ of the multiplets written in the original basis here in eq.~\eqref{eq:njl_vac}. Crucially $Q_5$ does not see the NJL interaction in its UV boundary conditions whilst $Q_{1,2,3}$ do. This will split the degeneracy of these four states, even though they share the same equation of motion, into a triplet and a singlet reflecting that the NJL term explicitly violates the SU(2) group with generators $T^{7,12,13}$. The triplet and singlet are characterised by their SU(2)$_V$ representations only. Note that for all the other scalar states and all the vector states we still require that they vanish in the UV of the bulk. 

Being careful to take into account the above issues, we can now compute the bound state spectrum as a function of $g_s^2$ using our solutions for $L_0(\rho)$ and $Q(\rho)$. We will name the states according to the basis eq.~\eqref{eq:njl_vac} and the fluctuations written in that basis as eq.~\eqref{fluc1}. We display the results for the scalar masses in figure \ref{fig:njl_flucs_vs_gssq} for a particular set of parameters listed in the caption.

The main features of the results for the scalar sector are as follows. For $g_s^2$ below the critical value only the $Q_{1,2,3,4}$ fields see the NJL interaction (through the boundary conditions on their masses in the bulk). These fields' masses fall to zero at the critical coupling as their effective potential readies to become unstable to a $Q$ vev above the critical coupling. 

Above the critical coupling $Q_{1,2,3}$ are true Goldstone bosons of the breaking of the axial SU(2) symmetry group. This can be seen directly since they obey eq.~\eqref{eq:sep_Q_eoms} with $L_1=L_p$ and $L_6=L_m$ and the $Q_{1,2,3}$ fields have Goldstone solutions take the form of the background $Q$ vev which correctly reproduces the UV value of $g_s^2$ of the background. $Q_4$, which is the sigma like field for the technicolor breaking pattern, grows sharply in mass above the critical $g_s^2$.

The remaining scalar states are all $g_s^2$ independent until above the critical coupling when they see the sharply rising $Q$ vev and their masses all grow. $\pi_{1,2,4}$ are a triplet that are accidentally degenerate here with the $\pi_5$ singlet. $\sigma$, $Q_5$ and the two $S_\pm$ states (made from mixtures of $S$ and $\pi_3$) are singlets. 

\begin{figure}
    \centering
    \includegraphics[width=0.8\textwidth]{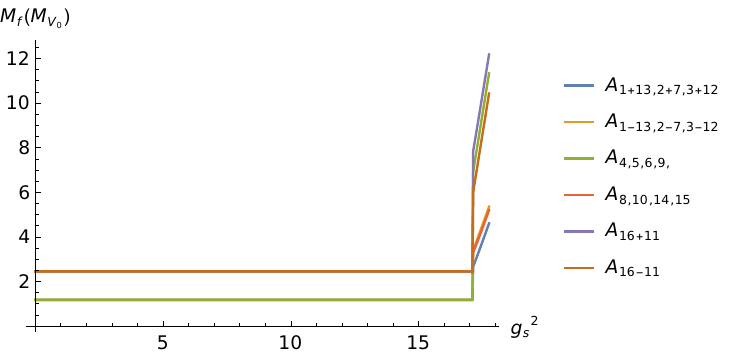} 
    \caption{Spin-1 bound states' masses. The multiplets are: two triplets associatd with generators $1 \pm 13, 2 \pm 7, 3 \pm12$; two sets of degenerate triplet and singlet associated with generators $4,5,6,9$ and $8,10,14,15$; and two singlets associated with generators $16 \pm 11$.}
    \label{fig:njl_spin_1_masses}
\end{figure}

 Now we turn to the vector mesons. The fields $V_{1+13,2+7,3+12}$ and 
$V_{1-13,2-7,3-12}$ transform as $\mathbf{3}$ of SU(2)$_V$ ($(\mathbf{1},\mathbf{3})$+$(\mathbf{3},\mathbf{1})$ of SU(2)$_L\times$ SU(2)$_R$), $A_{4,5,6,9}$ and $A_{8,10,14,15}$ transform both as $\mathbf{3}+\mathbf{1}$
of SU(2)$_V$ ($(\mathbf{2},\mathbf{2})$ of SU(2)$_L\times$ SU(2)$_R$) whereas $A_{16\pm 11}$ are both singlets of SU(2)$_V$ (SU(2)$_L\times$ SU(2)$_R$). In all cases the vector meson bulk wave functions vanish in the UV. We show the masses in figure~\ref{fig:njl_spin_1_masses}. Below the critical coupling they match the degenerate mass results we have previously described (splitting into the two groupings ``vector" and ``axial-vector" states). Above the critical coupling the additional symmetry breaking becomes apparent and the different multiplets split (with the exceptions of the singlets in the $(\mathbf{1},\mathbf{3})$+$(\mathbf{3},\mathbf{1})$ which remain accidentally degenerate). The overall signature though is that all the vector mesons' masses rise sharply above the critical coupling.

\begin{figure}
    \centering
    \begin{subfigure}{0.5\textwidth}
        \includegraphics[width=\textwidth]{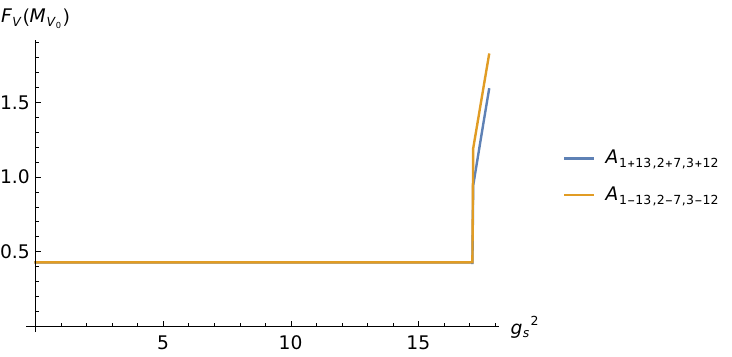}
        \caption{$f_{V}$}
    \end{subfigure}\hfill
    \begin{subfigure}{0.45\textwidth}
        \includegraphics[width=\textwidth]{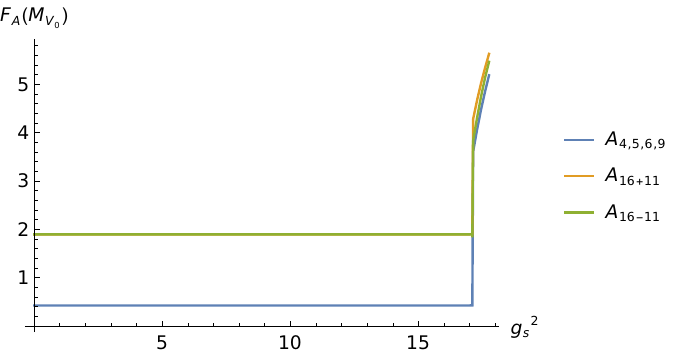}
        \caption{$f_{A}$}
    \end{subfigure}\hfill
    \begin{subfigure}{0.5\textwidth}
        \includegraphics[width=\textwidth]{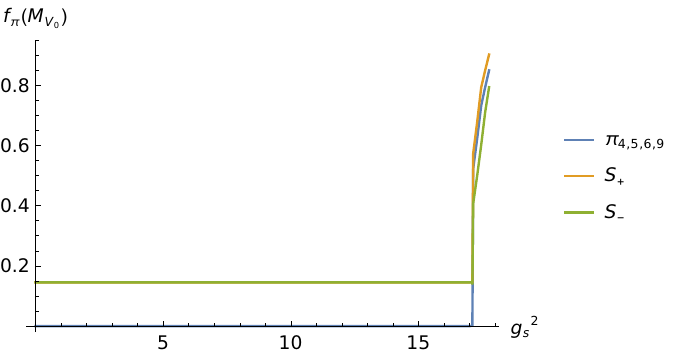}
        \caption{$f_{\pi}$}
    \end{subfigure}\hfill
        \begin{subfigure}{0.45\textwidth}
        \centering
        \includegraphics[width=\textwidth]{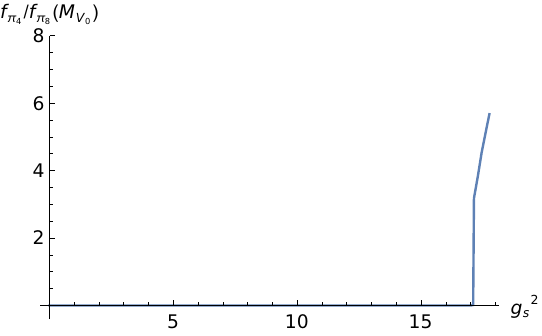}
        \caption{$f_{\pi_{4}}/f_{\pi_{8}}$}
        \label{fig:njl_f_vs_gssq_d}
    \end{subfigure}\hfill
    \caption{Meson decay constants vary with increasing scalar NJL coupling $g_s^2$, with gauge group SU(2), $B=0.1$.}
    \label{fig:njl_f_vs_gssq}
\end{figure}

 In figure~\ref{fig:njl_f_vs_gssq} we also show the vector decay constants. They show the same rise in scale and the same split into representations as the mass plots. Figure~\ref{fig:njl_f_vs_gssq_d}  shows how the electroweak breaking vev rises sharply relative to that of the underlying SU(2) theories vev scale. We again see that the rotation from composit Higgs to technicolour behaviour is very fine tuned in $g_s^2$ even with a relatively low UV scale for the NJL couplings. 

\section{Conclusions and outlook}

In this paper we have used a holographic model to study the spectrum of Sp(2N) gauge theories with four Weyl fermions in the fundamental representation. 

A key motivation for our work was to test the flexibility of holographic model building. We have had to incorporate the specific dynamics of each gauge theory through the running anomalous dimension of the fermion bilinear that condenses. We have worked with a non-abelian flavour structure that allows mass splittings and understanding of the bulk Higgsing of the gauge fields associated to the SU(4) flavour symmetry. Finally we have included four-fermion NJL operators to introduce a vacuum alignment competition between a ``composite Higgs" vacuum and a ``technicolour" vacuum. This has all been successfully achieved in a model where the inputs are the number of colours $N_c$, the strong-coupling scale of the gauge theory $\Lambda$, the masses of the fermions and the NJL coupling. The only caveat is that a further parameter $B$ was introduced that splits the $\pi$ and $Q$ scalar multiplets. The model is therefore very straightforwardly interpretable in the same parameter space as the true theory.  Holography is becoming a powerful tool for phenomenological studies of strongly coupled gauge theories.

Of course one must include the caveat that the model is just a model and not derived from first principles. The formalism does lie close in theory space to the known true duals of non-supersymmetrically deformed supersymmetric theories and our only change is essentially to the running anomalous dimensions in the theory. This is presumably why it gives sensible predictions even at a qualitative level. The only error estimates that can be made are by changing input parameters (we have computed the weak $N_c$ dependence of our results displaying to a degree this variability). Nevertheless, while we wait for expensive and time consuming lattice results the holographic model offers a quick rough and ready computational tool. 

In particular we have presented: the spectrum of the gauge theories at zero fermion mass in table~\ref{tab:deg_mass_tab}, the fermion mass dependence of the bound state masses in figure~\ref{fig:deg_flucs_vs_mq} ; and we have then looked at a particular example of a mass split theory in figure~\ref{fig:nondeg_flucs_vs_mq} . Finally in section~\ref{sec:njl} we have included NJL interactions and shown the fine tuned transition between the composite Higgs and technicolour vacuums that results, including following the spectrum through this transition. 

We hope our work will encourage further lattice work on these theories, be useful for beyond the standard model phenomenology and support dark matter model building.

\vspace{1cm}

\acknowledgments
Important elements of the work were carried out at the Mainz Institute for Theoretical Physics of the Cluster of Excellence PRISMA+ (Project ID 390831469) and NE and WP thank them for their hospitality.  We thank Giacomo Cacciapaglia and Kazem Bitaghsir Fadafan for in part motivating the work and discussions.
YL and WP are supported by the DFG,
project no.~PO-1337/11-1. NE's work was supported by the STFC consolidated grant ST/X000583/1.


\appendix
\section{Relation between $\dmq$ and $A$}
\label{app:dm_non_deg}
Here we work through the relation between $\dmq$ and $A$ as sketched in eq.~\eqref{eq:id_dm} for the full model including mass splitting and NJL interaction. The Lagrangian is 
\begin{equation}
    \label{eq:app_dm_nd_vac_lag}
    \begin{aligned}
        \mcl{L}&=\Tr\lrmb{\rho^3(\pdr L)^\dagger \pdr L+\rho A L^\dagger L}\\
        &=2\lrb{\rho^3(\pdr L_1)^2 +\rho^3(\pdr L_6)^2+\rho A L_1^2+\rho A L_6^2}.
    \end{aligned}
\end{equation}
Expanding in the fluctuations 
\begin{equation}
    \label{eq:app_dm_nd_lag}
    \begin{aligned}
       \half \mcl{L}&=\rho^3\lrb{\pdr (L_1+\d L_1)}^2+\lrb{\pdr (L_6+\d L_6)}^2\\
        &+\rho\lrb{A\bv{L_0}+{\pd A\over \pd L_1}\bv{L_0}\d L_1+{\pd A\over \pd L_6}\bv{L_0}\d L_6 +\half {\pd ^2A\over \pd L_1^2}\bv{L_0}\d L_1^2+\half {\pd ^2A\over \pd L_6^2}\bv{L_0}\d L_6^2+{\pd ^2A\over \pd L_1\pd L_6}\bv{L_0}\d L_1 \d L_6}\\
        &\times \lrmb{\lrb{L_1+\d L_1}^2+ \lrb{L_6+\d L_6}^2},
    \end{aligned}
\end{equation}
where $\bv{L_0}$ means evaluate the function at the vaccum. The equation of motion  of $\d L_1$ is 
\begin{equation}
    \label{eq:app_dm_L1_eom}
    \begin{aligned}
        &\pdbr{L_1}-\rho A\bv{L_0}L_1-{\rho\over2}\lrb{L_1^2+L_6^2}{\pd A\over \pd L_1} \bv{L_0} \\
        +&\pdbr{\d L_1}-\rho \lrmb{A\bv{L_0}+2{\pd A\over \pd L_1} \bv{L_0}L_1 +\half{\pd^2 A\over \pd L_1^2} \bv{L_0} \lrb{L_1^2+L_6^2}}\d L_1\\
        -&\rho\lrmb{{\pd A\over \pd L_6}\bv{L_0}L_1+2{\pd A\over \pd L_1}\bv{L_0}L_6+\half{\pd^2 A\over \pd L_1\pd L_6} \bv{L_0}\lrb{L_1^2+L_6^2}}\d L_6=0,
    \end{aligned}
\end{equation}
the equation of motion  for $\d L_6$ is similar, one just needs to exchange 1 with 6. From the two equations of motion, we define 
\begin{equation}
    \label{eq:app_nondeg_dm_id_full}
    \begin{aligned}
       A\bv{L_0}L_1+{1\over2}\lrb{L_1^2+L_6^2}{\pd A\over \pd L_1} \bv{L_0}&\equiv \dmq\lrmb{\log\sqrt{{1\over4}\Tr\lrb{\rho^2+L^\dagger L}}} L_1,   \\
        A\bv{L_0}L_6+{1\over2}\lrb{L_1^2+L_6^2}{\pd A\over \pd L_6} \bv{L_0}&\equiv \dmq\lrmb{\log\sqrt{{1\over4}\Tr\lrb{\rho^2+L^\dagger L}}} L_6.
    \end{aligned}
\end{equation}
Multiply both equations by $L_1$ and $L_6$, respectively, and adding them together, this is nothing but
\begin{equation}
    \label{eq:app_nondeg_dm_id}
       A\bv{L_0}+2\lrb{L_1^2+L_6^2}{d A\over d \text{Arg}} \bv{L_0}\equiv \dmq\lrmb{\log\sqrt{{1\over4}\Tr\lrb{\rho^2+L^\dagger L}}},
\end{equation}
where $\text{Arg}=\Tr(\rho^2\mathbbm{1}_4+ X^\dagger X$), ${d A\over d \text{Arg}}\equiv A'$, and
\begin{equation}
    \label{eq:app_rarg}
    {1\over 4}\Tr(\rho^2\mathbbm{1}_4+L^\dagger L)=\rho^2+\half(L_1^2+L_6^2)\equiv r_{arg,16}^2.
\end{equation}
Expanding the above equation in fluctuations to first order
\begin{equation}
    \label{eq:app_derive_nondeg_dmp}
    \begin{aligned}
        &A\bv{L_0}+{\pd A\over \pd L_1}\bv{L_0}\d L_1+{\pd A\over \pd L_6}\bv{L_0}\d L_6+2\lrmb{\lrb{L_1+\d L_1}^2+\lrb{L_6+\d L_6}^2}\lrb{A'\bv{L_0}+{\pd A'\over\pd L_1}\bv{L_0}\d L_1+{\pd A'\over\pd L_6}\bv{L_0}\d L_6}\\   =&A\bv{L_0}+2(L_1^2+L_6^2)A'\bv{L_0}\rightarrow\dmq\bv{L_0}\\
        +&\lrb{{\pd A\over \pd L_1}\bv{L_0}+4L_1 A'\bv{L_0}+2\lrb{L_1^2+L_6^2}{\pd A'\over \pd L_1}\bv{L_0}}\d L_1\rightarrow {\pd \dmq\over\pd L_1}\d L_1\\
        +&\lrb{{\pd A\over \pd L_6}\bv{L_0}+4L_6 A'\bv{L_0}+2\lrb{L_1^2+L_6^2}{\pd A'\over \pd L_6}\bv{L_0}}\d L_6\rightarrow {\pd \dmq\over\pd L_6}\d L_6.\\
    \end{aligned}
\end{equation}
Notice that 
\begin{equation}
    \label{eq:app_aux_ddm}
    \begin{aligned}
     &{\pd \dmq\over \pd L_1}=\half{L_1\over \rho^2+\half(L_1^2+L_6^2)}{d \dmq\over d (\log r_{arg,16})}\equiv\half{L_1\over \rho^2+\half(L_1^2+L_6^2)}\dmqp,   \\
    &{\pd \dmq\over \pd L_6}=\half{L_6\over \rho^2+\half(L_1^2+L_6^2)}\dmqp,
    \end{aligned}
\end{equation}
we find 
\begin{equation}
    \label{eq:app_ddm_id}
    8A'+8\lrb{L_1^2+L_6^2}A''=\half{1\over \rho^2+\half\lrb{L_1^2+L_6^2}}\dmqp,\quad A''={d^2A\over d \text{Arg}^2 }.
\end{equation}
Using the above relations, we get the equation of motions as in \cref{sec:non_degenerate_case}. 

The same computation is done for the NJL case. Without repeating the details, we only list the result here
\begin{equation}
    \label{eq:app_ddm_njl_id}
    16\lrb{A'+\lrb{L_p^2+L_m^2}A''}={1\over \rho^2+\half\lrb{L_p^2+L_m^2}}\dmqp.
\end{equation}

\bibliographystyle{utphys}
\bibliography{main}

\end{document}